\documentclass[preprint,12pt]{elsarticle}

\usepackage{color}
\usepackage{amssymb}

\journal{Astroparticle Physics}

\begin{document}

\begin{frontmatter}



\title{A Detailed Study on the Range Fluctuations of $10^{11}$eV to $10^{18}$eV Muons in Water and the Fluctuations of the Cherenkov Lights due to the Accompanied Cascade Showers initiated by Muons}


\author[1]{Y.Okumura}
\author[1]{N.Takahashi}
\author[2]{A.Misaki}

\address[1]{Graduate School of Science and Technology Hirosaki University, Hirosaki, 036-8561, Japan}
\address[2]{Innovative Research Organization, Saitama University, 338-8570, Japan}

\begin{abstract}
 The validity of our Monte Carlo simulation procedure ({\it the integral method}) had been verified by the corresponding analytical procedure of which is quite independent of our method methodologically. Also, the results obtained by our procedure are compared with those  obtained by the different Monte Carlo simulation procedure ({\it the differential method}) which have been exclusively utilized by the different authors and the agreement between them are found to be well. By utilizing our Monte Carlo procedures, the validity of which is guaranteed in two different procedures, we investigate not only  the fluctuation of high energy muons themselves but also fluctuation of the various quantities related to the energy losses  by the muons, which are difficult to obtain by {\it the differential method}. Namely, we obtain fluctuation on energy losses of the muons, fluctuation on Cherenkov lights due to the accompanied cascade showers initiated by the muon and the correlations between them .
Finally, we obtain the transition curves for Cherenkov lights in KM3 detector, taking into account of all possible fluctuations in the stochastic processes and point out the difficulty of the reliable estimation of the energy of the muons which are resultants of muon neutrino events in the KM3 detectors.
\end{abstract}


\end{frontmatter}


\section{Introduction}
\label{intro}
The fluctuation in high energy muon's behavior may play an important role in the analysis of muon 
neutrino events  for KM3 detector deployed in the Antarctic, the ocean and the lake \cite{IceCube} \cite{Antarctic} \cite{Mediterranean} \cite{Baikal}. 
 As far as the treatment of the range fluctuation of high energy muons by the 
Monte-Carlo method is concerned, there exist two independent 
approaches. The one is {\it the differential method} in which 
the muons concerned are pursued in step by step way 
\cite{Lipari} \cite{Antonioli} \cite{Dutta} \cite{Klimushin} \cite{Chirkin} \cite{Kudryavtsev} \cite{Bottai}.
In this method, the quantity of $v_{cut}$ is introduced so as to separate the continuous parts from the radiative parts in the stochastic processes in order to save the time for computation. 
 The other is {\it the integral method} in which the interaction points of the muons and their dissipated energy are directly determined \cite{Takahashi}\cite{Takahashi2} and here, all the processes are treated in the stochastic manner without the introduction of $v_{cut}$.
These two methods are independent form each other, but are logically equivalent, giving the same results as for the muons' behaviors (see Figures 4 to 6). 
 However, it should be noticed that the energy determination of the high energy muons made by the measurement of the Cherenkov lights which are produced by the accompanied cascade showers. 
These cascade showers are generated from the stochastic processes, such as bremsstrahlung, direct electron pair production and photo nuclear nteraction which are initiated by muons concerned. 


\section{Range fluctuation of the (ultra-) high energy 
muons and individual behavior of the muons}
\label{sec:2}
\subsection{The physical meaning of "no fluctuation"}
\label{sec:2.1}
\begin{figure}[h]
\begin{center}
\begin{tabular}{cc}
\begin{minipage}{0.5\hsize}
\resizebox{0.9\textwidth}{!}{
\includegraphics{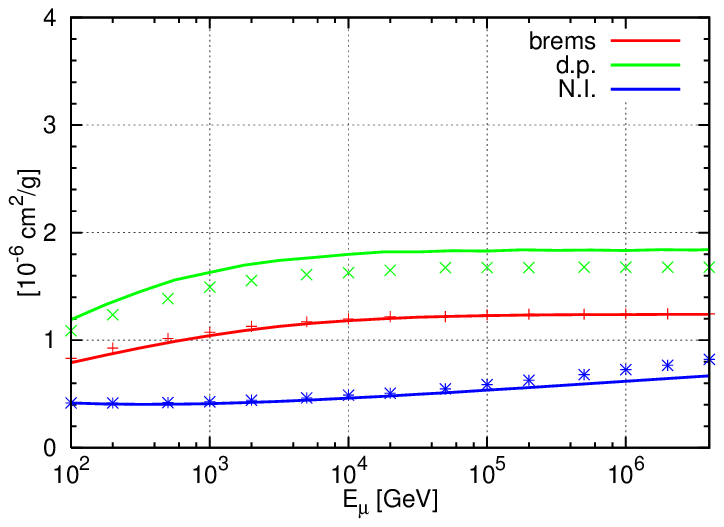}}		
\caption{b-terms due to different stochastic processes}
\label{fig:B-TERM_W}
\end{minipage}
\begin{minipage}{0.5\hsize}
\resizebox{0.9\textwidth}{!}{
\includegraphics{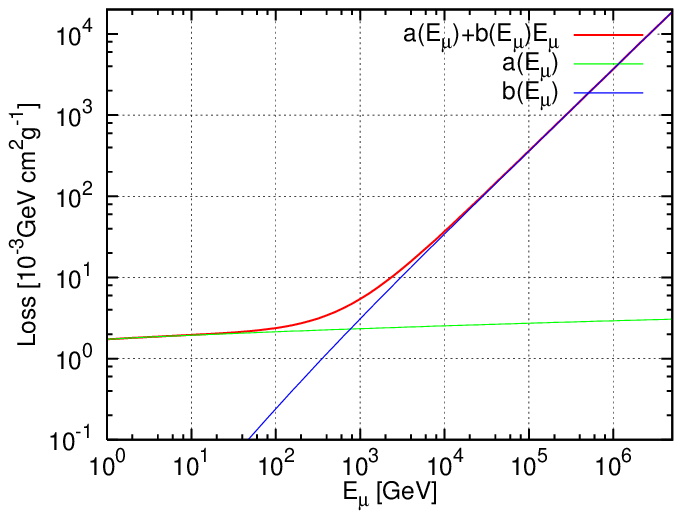}}
\caption{The relation between {\it a}-term and {\it b}-term in water}
\label{fig:dEdx}	                       
\end{minipage}
\end{tabular}
\end{center}
\end{figure}

\ The average energy loss by high energy muon is usually 
described as,
\begin{equation}
\label{dEdx}	
\frac{dE_{\mu}}{dx}={\it a}\left(E_{\mu}\right)+{\it b}\left(E_{\mu}\right)\cdot E_{\mu},
\end{equation}
where {\it a} is the term due to ionization which is free from 
fluctuation and {\it b} is the term due to stochastic processes 
which may be origins of fluctuations. The latter is divided 
into three parts. Namely,
\begin{equation}
\label{bEmu}	
{\it b}\left(E_{\mu}\right)={\it b}_{brems}\left(E_{\mu}\right)+{\it b}_{d.p}\left(E_{\mu}\right)+{\it b}_{nucl}\left(E_{\mu}\right),
\end{equation}
where ${\it b}_{brems}$, ${\it b}_{d.p}$ and ${\it b}_{nucl}$ are the corresponding terms due to bremsstrahlung, 
direct electron pair production and photo nuclear interaction, respectively.
 In the treatment of the average energy loss, each {\it b} 
term is defined as,
\begin{equation}
\label{intb}	
{\it b}_{i}\left(E_{\mu}\right)=\frac{N}{A}\int_{v_{min}}^{v_{max}}v\left[\frac{d\sigma_{i}\left(v,E_{\mu}\right)}{dv}\right]\cdot dv,
\end{equation}
where $v_{max}$ and $v_{min}$ are the maximum and the minimum 
fractional energies due to their kinematical limits.
The physical meaning of Eq.(\ref{dEdx}) is that the muons concerned disipate energy uniquely, being defined by Eq.(\ref{bEmu}), namely, fluctuations are not included in Eq.(\ref{bEmu}).
\\
 In Figure \ref{fig:B-TERM_W}, we give the {\it b} terms due to different 
processes in water.
\\
 In Figure \ref{fig:dEdx}, we give the relation between $a(E_{\mu})$ 
and $b(E_{\mu})$. As the b-terms essentially are of stochastic character, it is seen from
the figure that the stochastic 
processes become effective above $\sim$ 1 TeV. Therefore, we must 
treat the muon's behavior in stochastic manner above $\sim$ 1 
TeV. Below $\sim$ 1 TeV we may treat muon's behavior in the 
non-stochastic manner.\\
Then, the range of the muon is uniquely determined by Eq.(\ref{Rdx}).
\begin{equation}
\label{Rdx}	
R=\int_{E_{min}}^{E_{0}}\frac{dx}{-\frac{dE_{\mu}}{dx}},
\end{equation}
where $E_{min}$ is the minimum energy for observation and 
$E_{0}$ is the primary energy of the muon.
\ Through the present paper $E_{min}$ denotes the minimum 
energy $E_{min}$ among the energies for observation ($E_{obs}$) and it is 
taken as 1 GeV. Thus, $R$ defined by Eq.(\ref{Rdx}) gives the effective 
range of the muon without fluctuation. Exactly speaking, $R$ is 
the muon range where the fluctuation effects in the stochastic 
processes are neglected.
The physical meaning of "no fluctuation" is that the muons in the stochastic processes lose their energies in the form of the effective energy loss defined by Eq.(\ref{intb}).

\begin{figure}[t]
\begin{center}
\resizebox{0.7\textwidth}{!}{
\includegraphics{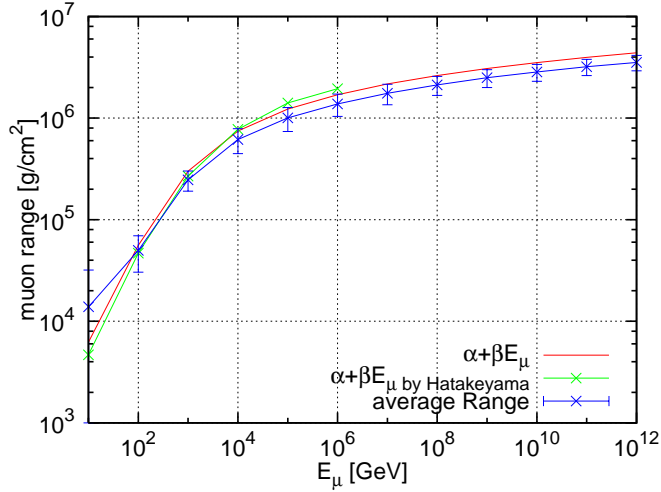}}
\caption{The range energy relation between the case without fluctuation and the case with fluctuation in standard rock. The uncertainty bar denote the standard deviation.}
\label{fig:RangeSR}	            
\end{center}
\end{figure}

 In Figure \ref{fig:RangeSR}, we give the effective range defined by Eq.(\ref{Rdx}) together with the average ranges of the muons in which the fluctuation effects are exactly taken into account.(see, discussion in the later sections). 

 It should be noticed from the figure that the effective ranges without fluctuation are different from the average range of the muons in which the fluctuation is considered, which Lipari and Stanev\cite{Lipari} already pointed out. Really, the real average ranges are smaller than those of effective range beyond one standard deviation above $10^{13}$ eV as shown Figure \ref{fig:RangeSR}.
\subsection{Physical quantities with fluctuation}
\label{sec:2.2}
\ In {\it the differential method}, many authors \cite{Lipari},\cite{Antonioli},\cite{Dutta},\cite{Klimushin},\cite{Chirkin},\cite{Kudryavtsev},\cite{Bottai} divide all stochastic processes 
into two part, namely, the continuous part and radiative
part in their Monte Carlo simulation in order 
to consider the fluctuation in both the range and energies of the muons and introduce $v_{cut}$ to save time for computation, while we treat 
all stochastic processes as exactly as possible without the 
introduction of $v_{cut}$. The validity of our Monte Carlo method had been checked by the corresponding analytical method which is methodologically independent of the Monte Carlo procedure and the essential structure of our Monte Carlo method is described in the Appendix. Furthermore, we check the validity of our method, comparing our results with the corresponding result by \textit{the differential methods}, which are shown in (\ref{sec:2.2.1}).
\subsubsection{The comparison of our results with others}
\label{sec:2.2.1}
Our survival probability for high energy muon is defined as,
\begin{equation}
\label{PEXE}	
P\left(E_{obs},X,E_{p}\right)=\frac{N_{through}\left(E_{obs},X\right)}{N_{sample}\left(E_{p}\right)},
\end{equation}
where $E_{p}$, $X$, and $E_{obs}$, denote the energy of primary muon, the point for observation and the minimum energy of the muon at the point for observation, respectively.   
$N_{sample}\left(E_{p}\right)$ denotes the total sampling number of muons and $N_{through}\left(E_{obs},X\right)$ denotes the number of muons concerned with energies above $E_{obs}$ which pass through the observation point X.
\begin{figure}[b]
\begin{center}
\begin{tabular}{ccc}
\begin{minipage}{0.33\hsize}
\resizebox{1.0\textwidth}{!}{
\includegraphics{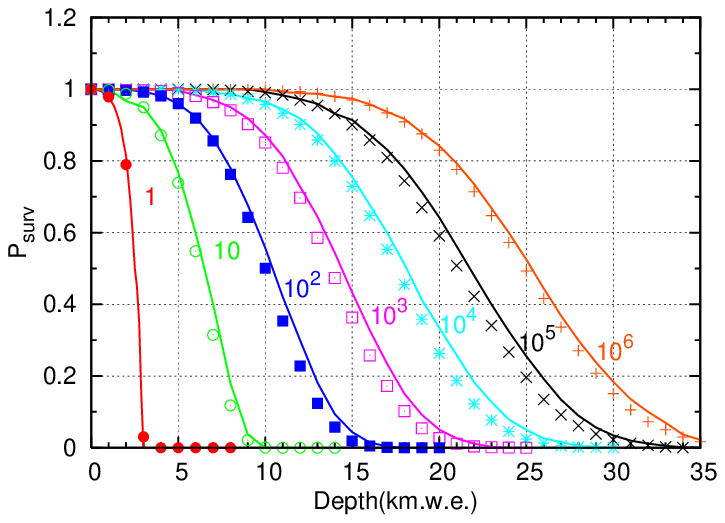}}		
\caption{The comparison of our result with that of Lipari and Stanev}
\label{fig:Lipari}
\end{minipage}
\begin{minipage}{0.33\hsize}
\resizebox{1.0\textwidth}{!}{
\includegraphics{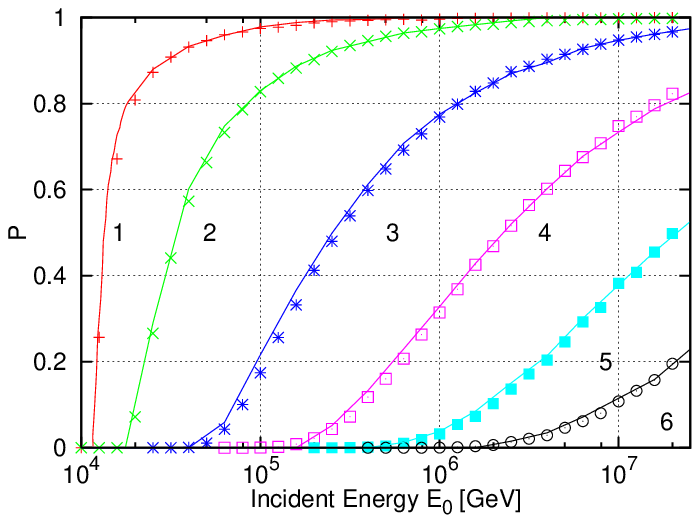}}	
\caption{The comparison of our results with that of Klimushin et al}
\label{fig:Buga}
\end{minipage}
\begin{minipage}{0.33\hsize}
\resizebox{1.0\textwidth}{!}{
\includegraphics{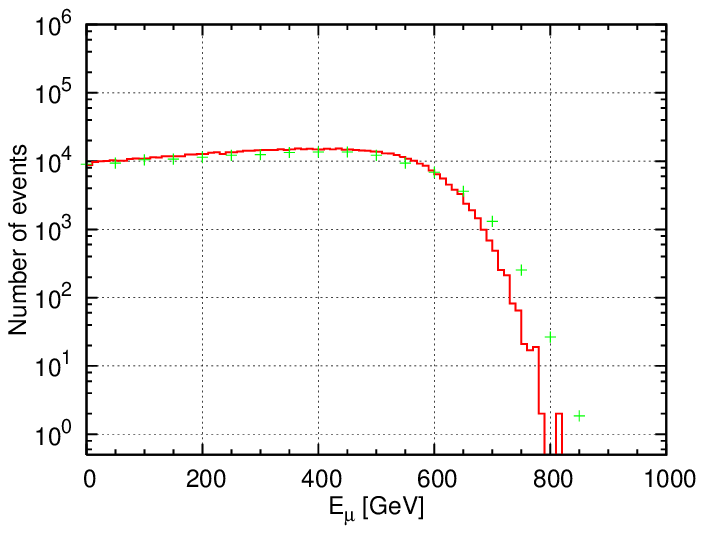}}	
\caption{The comparison of our results with that of Kudryavtsev}
\label{fig:ES3km}
\end{minipage}
\end{tabular}
\end{center}
\end{figure}

We compare our results by {\it the integral method} with the different authors' results by {\it the differential method} in the following.
Lipari and Stanev\cite{Lipari} give the survival probabilities as the functions of the depths for $10^{11}$eV to $10^{18}$eV  incident muons in water and partly standard rock, the minimum energy of which is taken as 1 GeV. We obtain the corresponding results by {\it the integral method} and compare our results with Lipari and Stanev's in Figure \ref{fig:Lipari}. Also, Klimushin et al\cite{Klimushin} give the survival probabilities for primary energy of $10^{13}$eV to $3 \times 10^{16}$eV. We obtain the corresponding results to them and compare our corresponding results with the results by Klimushin et al\cite{Klimushin} in Figure \ref{fig:Buga}. Furthermore, Kudryavtsev\cite{Kudryavtsev} gives the energy spectrum of the muon due to primary energy of 2 TeV at 3 km in water. We obtain the corresponding results to him and compare our corresponding results with his results in Figure \ref{fig:ES3km}. The agreements between the different authors' result obtained by {\it the differential method} and our results obtained by {\it the integral method} are well as shown in Figures 3 to 5, taking into account of the slight differences in the cross sections utilized between the different authors' and ours. It is seen from these figures that the validity of our {\it integral method} is guaranteed by {\it the differential method} due to the different authors.
\begin{figure}[b]
\begin{center}
\begin{tabular}{ccc}
\begin{minipage}{0.33\hsize}
\resizebox{1.0\textwidth}{!}{
\includegraphics{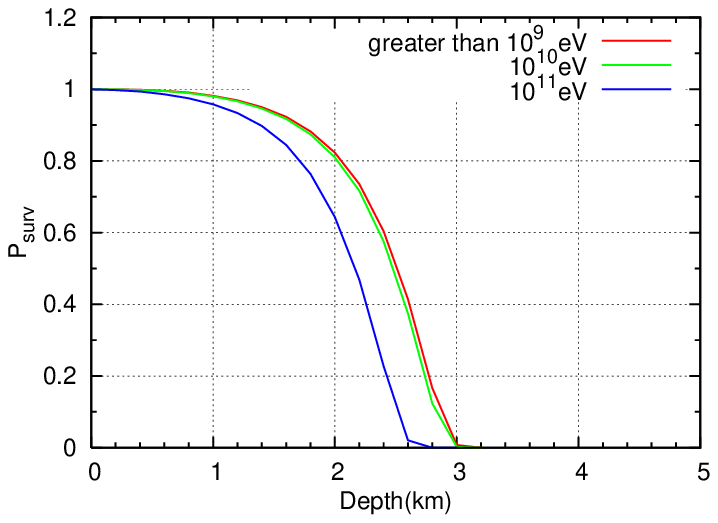}}	        
\caption{The survival probabilities for $10^{12}$eV muon}
\label{fig:SP12}
\end{minipage}
\begin{minipage}{0.33\hsize}
\resizebox{1.0\textwidth}{!}{
\includegraphics{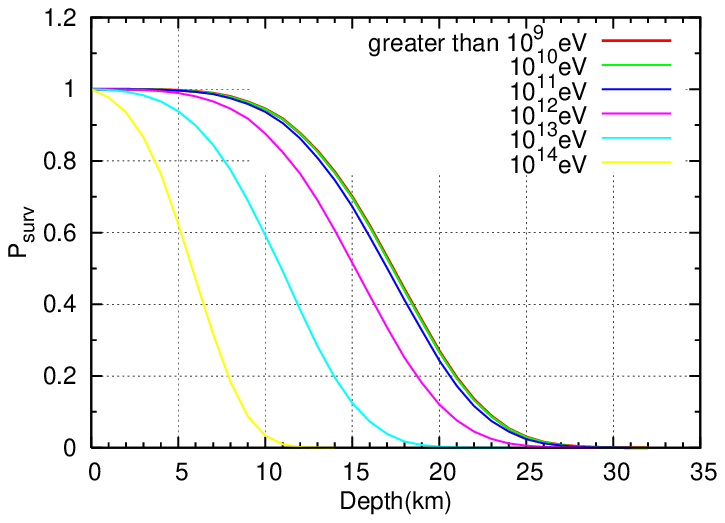}}	        
\caption{The survival probabilities for $10^{15}$eV muon}
\label{fig:SP15}
\end{minipage}
\begin{minipage}{0.33\hsize}
\resizebox{1.0\textwidth}{!}{
\includegraphics{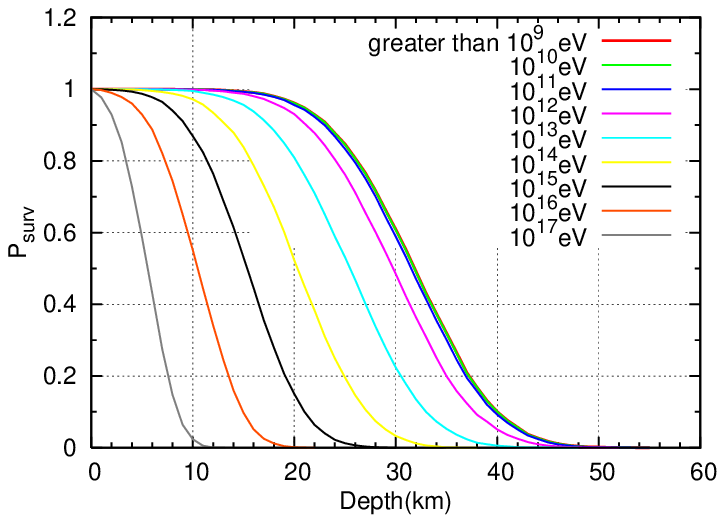}}	        
\caption{The survival probability for $10^{18}$eV}
\label{fig:SP18}
\end{minipage}
\end{tabular}
\end{center}
\end{figure}

As already explained, all the authors due to {\it the differential method} whose results are compared with ours divide the stochastic processes into two parts, namely, the radiative part and the continuous part to perform the Monte Carlo calculation for the study on the fluctuation of the muon behaviors.  For the purpose, they introduce $v_{cut}$ by which they separate the radiation processes from the continuous part and they study the fluctuation effect of the muon in the radiative part only.

\begin{eqnarray}
\label{dEdx_rad}  
\left[\frac{dE}{dx}\right]_{rad}&=&\left[\frac{dE}{dx}\right]_{soft}+\left[\frac{dE}{dx}\right]_{rad} \nonumber \\
&=&\frac{N}{A}E\int_{0}^{v_{cut}}dv\cdot v\frac{\sigma\left(v,E\right)}{dv}+\frac{N}{A}E\int_{v_{cut}}^{1}dv\cdot v\frac{d\sigma\left(v,E\right)}{dv}
\end{eqnarray}
 
 Such treatment is logically correct only as far as we are interested in the muon behaviors, because the energy loss by the muon with single primary energy is exactly taken into account in their treatment irrespective of any $v_{cut}$. 
However, if we are interested in Cherenkov radiation responsible for all stochastic processes through the accompanied cascade showers, then, the methods adopted by these authors are not adequate for the study on such the purpose, on which we discuss in later section. (see, section \ref{sec:4})
\subsubsection{Survival probabilities and  their differential energy spectra at different observable depth}
\label{sec:2.2.2}
 In Figures \ref{fig:SP12} to \ref{fig:SP18}, we give the survival probabilities for different observable energies with primary energies of $10^{12}$, $10^{15}$ and $10^{18}$eV, respectively. 
 In Figure \ref{fig:SP12}, we give the minimum observation energies $10^{9}$eV, $10^{10}$eV and $10^{11}$eV, respectively. In Figure \ref{fig:SP15}, we give them $10^{9}$eV, $10^{10}$eV, $10^{11}$eV, $10^{12}$eV, $10^{13}$eV and $10^{14}$eV, respectively. In Figure \ref{fig:SP18}, we give them, $10^{9}$eV to $10^{17}$eV, respectively. Each sampling number in Figure 7 to 9 is 100,000. It is seen from the figures that the survival probabilities become remarkably large as their primary energies increase.

\begin{figure}[h]
\begin{center}
\begin{tabular}{ccc}
\begin{minipage}{0.33\hsize}
\resizebox{1.0\textwidth}{!}{
\includegraphics{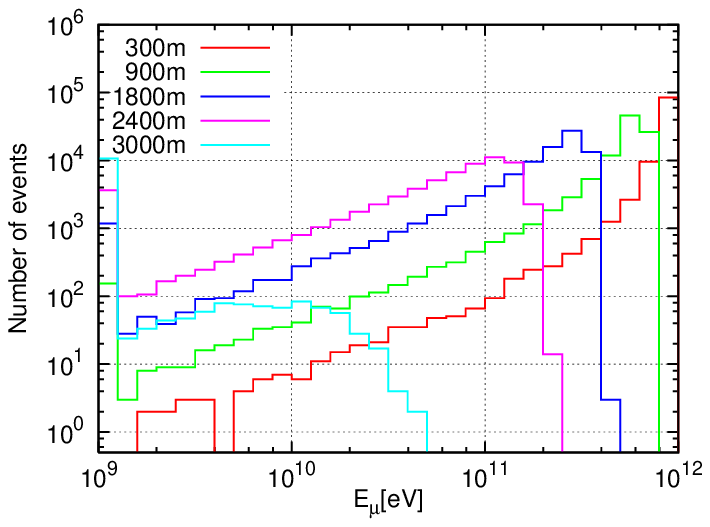}}	
\caption{Energy spectrum at the different depths, initiated by $10^{12}$eV muons}
\label{fig:ES12}
\end{minipage}
\begin{minipage}{0.33\hsize}
\resizebox{1.0\textwidth}{!}{
\includegraphics{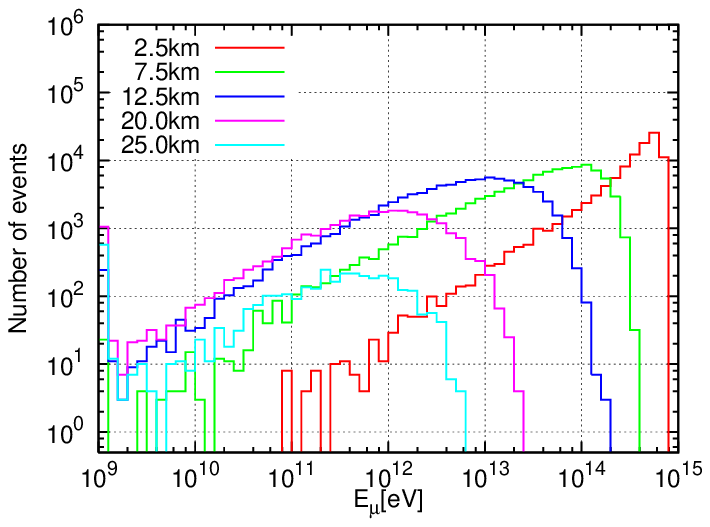}}	
\caption{Energy spectrum at the different depths, initiated by $10^{15}$eV muons}
\label{fig:ES15}
\end{minipage}
\begin{minipage}{0.33\hsize}
\resizebox{1.0\textwidth}{!}{
\includegraphics{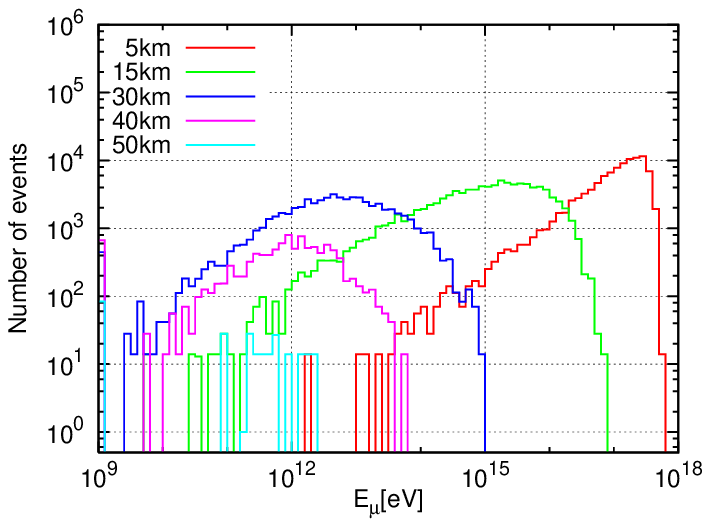}}	
\caption{Energy spectrum at the different depths, initiated by $10^{18}$eV muons}
\label{fig:ES18}
\end{minipage}
\end{tabular}
\end{center}
\end{figure}

 In Figures \ref{fig:ES12} to \ref{fig:ES18}, we give the differential energy spectra for primary energies, $10^{12}$, $10^{15}$ and $10^{18}$eV at different depths, respectively.
\textcolor{black}{It is seen from the figures that the energy spectrum at the initial stage are of delta-function and they shift as the mountain-like deforming their shape in the intermediate stage and finally, they disappear as the results of the delta-function type again. Each sampling number in Figures 10 to 12 is 100,000.}
\subsubsection{Range Distribution of Muon}
\begin{figure}[h]
\begin{center}
\resizebox{0.7\textwidth}{!}{
\includegraphics{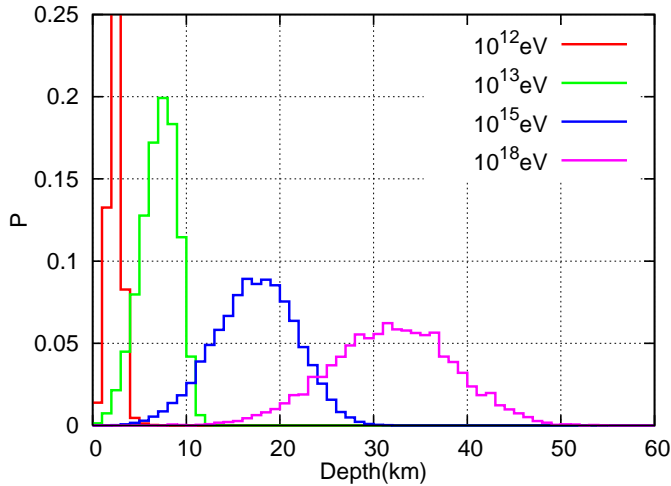}}	        
\caption{Range distributions for $10^{12}$eV to $10^{18}$eV muons. The minimum observation energies are taken as $10^{9}$ eV. Each sampling number is 100,000.}
\label{fig:RF131518}
\end{center}
\end{figure}
\ All processes, such as bremsstrahlung, direct electron pair production and photo nuclear interaction are of stochastic ones and, therefore, one cannot neglect their fluctuation essentially. The muons propagate through the matter as the results of the competition effect among bremsstrahlung, direct electron pair production and photo nuclear interaction.

In Figure \ref{fig:RF131518}, we give $P\left(R;E_{p}\right)$, the probabilities for the range distribution with primary energies, $10^{13}$, $10^{15}$ and $10^{18}$eV, respectively. It is clear from the figures that the width of the range distribution increases rapidly, as their primary energy increases. Also, as the primary energy decreases, the width of range distribution becomes narrower and approaches to the delta function-type, the limit of which denotes no fluctuation. 
Their average ranges and their standard deviations are given Table \ref{tab:AVE}.
\begin{table}[h]        
\begin{center}
\caption{The average values and values of their standard deviations of the range distributions of muons which are expressed as the normal distribution. These numerical values are given in both the km and $g/cm^{2}$ for primary energies from $10^{11}$eV to $10^{18}$eV.}
\label{tab:AVE}
\scalebox{1.0}[1.0]{
\begin{tabular}{c|c|c|c|c}
\hline\noalign{\smallskip}
$E_{p}$&$<R>$&$<R>$&$\sigma$&$\sigma$\\
$[eV]$&$[g/cm^{2}]$&$[km]$&$[g/cm^{2}]$&$[km]$\\
\noalign{\smallskip}\hline
$10^{11}$&4.75$\times 10^{4}$&4.75$\times 10^{-1}$&2.39$\times 10^{4}$&2.39$\times 10^{-1}$\\ \hline
$10^{12}$&2.51$\times 10^{5}$&2.51$\times 10^{0}$&5.32$\times 10^{4}$&5.32$\times 10^{-1}$\\ \hline
$10^{13}$&7.18$\times 10^{5}$&7.18$\times 10^{0}$&1.97$\times 10^{5}$&1.97$\times 10^{0}$\\ \hline
$10^{14}$&1.23$\times 10^{6}$&1.23$\times 10^{1}$&3.35$\times 10^{5}$&3.35$\times 10^{0}$\\ \hline
$10^{15}$&1.73$\times 10^{6}$&1.73$\times 10^{1}$&4.35$\times 10^{5}$&4.35$\times 10^{0}$\\ \hline
$10^{16}$&2.22$\times 10^{6}$&2.22$\times 10^{1}$&5.16$\times 10^{5}$&5.16$\times 10^{0}$\\ \hline
$10^{17}$&2.70$\times 10^{6}$&2.70$\times 10^{1}$&5.86$\times 10^{5}$&5.86$\times 10^{0}$\\ \hline
$10^{18}$&3.19$\times 10^{6}$&3.19$\times 10^{1}$&6.52$\times 10^{5}$&6.52$\times 10^{0}$\\ \hline\noalign{\smallskip}
\end{tabular}
}
\end{center}
\end{table}
 Then, the range distribution can be well approximated as the normal distribution in the following.
\begin{equation}
\label{PRE}  
P\left(R;E_{p}\right)=\frac{1}{\sqrt{2\pi}\sigma}exp\left(-\frac{R-<R>}{2\sigma^{2}}\right),
\end{equation}
where $E_{p}$, $<R>$ and $\sigma$ are primary energy, the average value of ranges and the standard deviations, respectively.

In order to examine the nature of the stochastic processes in each process further, we compare the real range distribution in which each stochastic process (bremsstrahlung, direct electron pair production and photo nuclear interaction) are taken into account as the competition effect with the hypothetical range of the muon due to each stochastic process.

\begin{figure}[b]
\begin{center}
\begin{tabular}{ccc}
\begin{minipage}{0.33\hsize}
\resizebox{1.0\textwidth}{!}{
\includegraphics{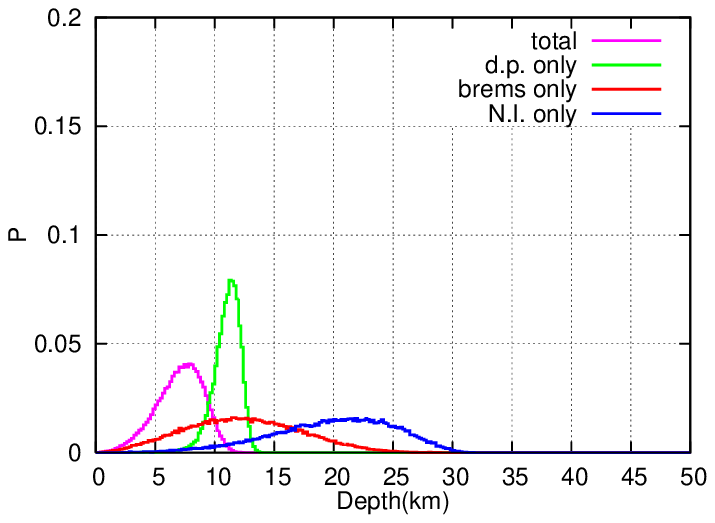}}	
\caption{Hypothetical range distributions for $10^{13}$eV muons.}
\label{fig:RF13Sp}
\end{minipage}
\begin{minipage}{0.33\hsize}
\resizebox{1.0\textwidth}{!}{
\includegraphics{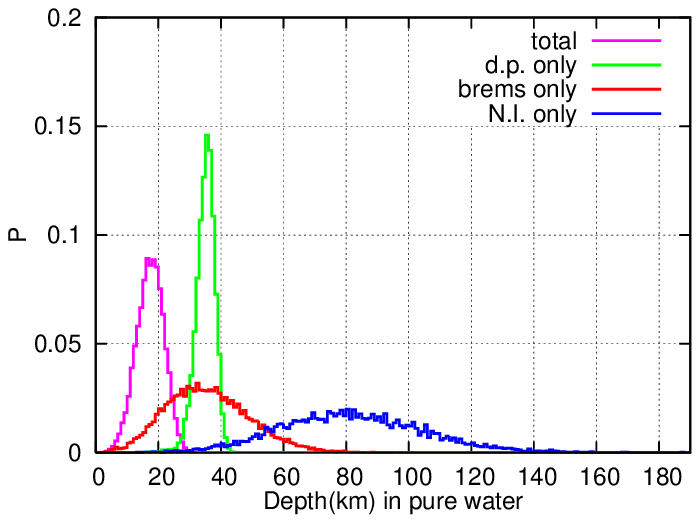}}	
\caption{Hypothetical range distributions for $10^{15}$eV muons.}
\label{fig:RF15Sp}
\end{minipage}
\begin{minipage}{0.33\hsize}
\resizebox{1.0\textwidth}{!}{
\includegraphics{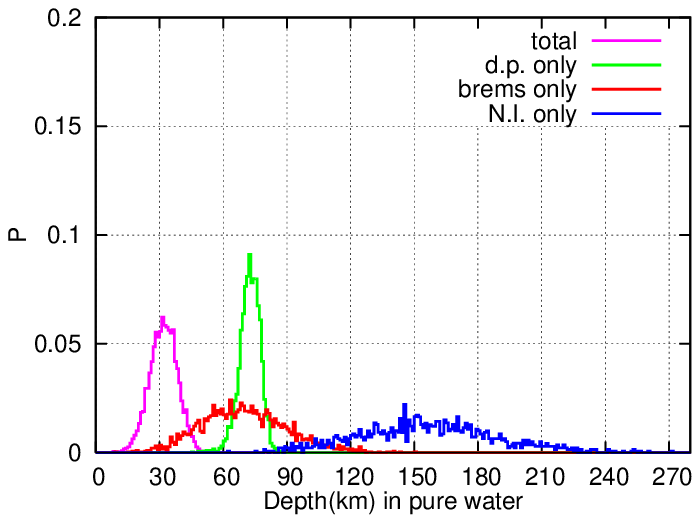}}	
\caption{Hypothetical range distributions for $10^{18}$eV muons.}
\label{fig:RF18Sp}
\end{minipage}
\end{tabular}
\end{center}
\end{figure}

In addition to the real range distributions, we give the hypothetical range distributions for $10^{13}$, $10^{15}$ and $10^{18}$ eV, in which only one cause among three stochastic processes is taken into account as shown in Figures \ref{fig:RF13Sp} to \ref{fig:RF18Sp}.
 Here, the symbol of [d.p.only] in these figures means the hypothetical range distribution in which only direct electron pair production is taken into account and the bremsstrahlung and nuclear interaction are neglected. The symbols of [brems only] and [N.I. only] have similar meaning to that of [d.p.only].
 From the shapes of the distribution and their maximum frequencies for different stochastic processes in Figures \ref{fig:RF13Sp} to \ref{fig:RF18Sp}, it is clear that energy loss in the direct electron  pair production is of small fluctuation, while both the bremsstrahlung and photo-nuclear interaction are of bigger fluctuation and the fluctuation in photo nuclear interaction becomes remarkable when compared with bremsstrahlung as primary energy increases.
\subsubsection{The diversity of individual muon behavior}
In Figures 17 to 25, we show the diversities of the muons' behaviors for the same primary energy of muons with regard to their ranges (or their energy losses).
\begin{figure}[h]
\begin{center}
\begin{tabular}{ccc}
\begin{minipage}{0.33\hsize}
\resizebox{1.0\textwidth}{!}{
\includegraphics{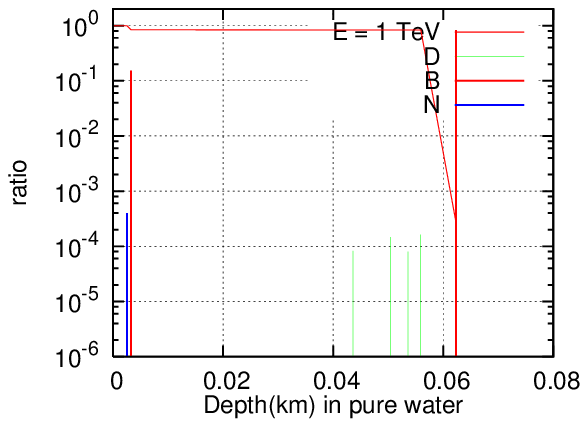}}	
\caption{The energy losses with the shortest range for $10^{12}$eV muons.}
\label{fig:ELS12}
\end{minipage}
\begin{minipage}{0.33\hsize}
\resizebox{1.0\textwidth}{!}{
\includegraphics{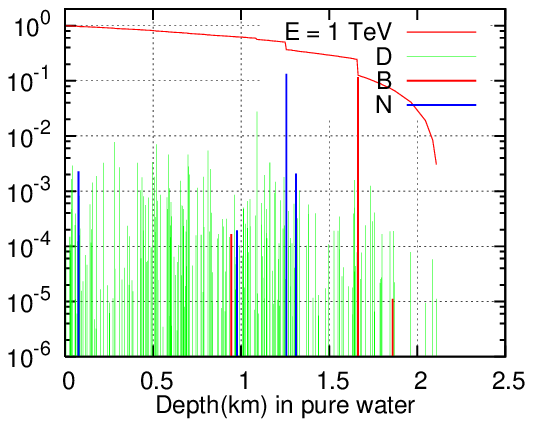}}	
\caption{The energy losses with the average-like range for $10^{12}$eV muons.}
\label{fig:ELA12}
\end{minipage}
\begin{minipage}{0.33\hsize}
\resizebox{1.0\textwidth}{!}{
\includegraphics{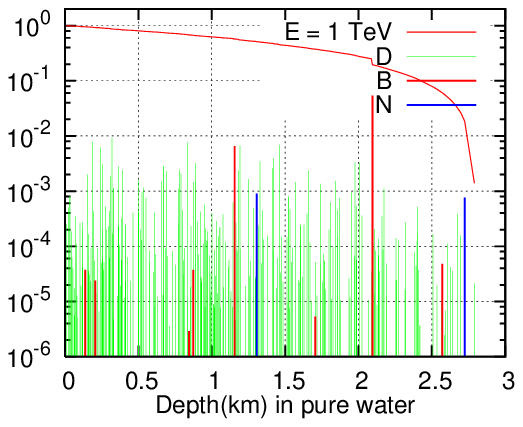}}	
\caption{The energy losses with the longest range for $10^{12}$ eV muons }
\label{fig:ELL12}
\end{minipage}
\end{tabular}
\end{center}
\end{figure}
\begin{figure}[h]
\begin{center}
\begin{tabular}{ccc}
\begin{minipage}{0.33\hsize}
\resizebox{1.0\textwidth}{!}{
\includegraphics{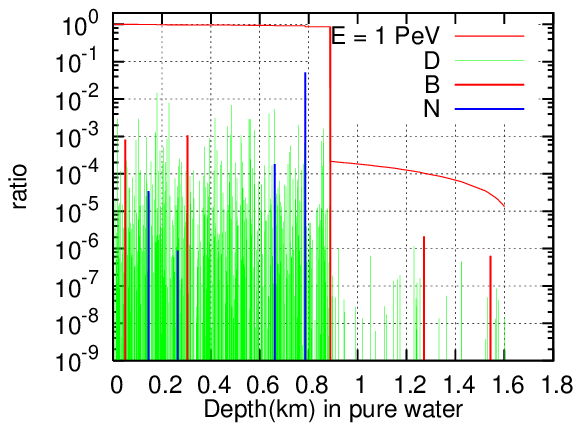}}	
\caption{The energy losses with the shortest range for $10^{15}$eV muons.}
\label{fig:ELS15}
\end{minipage}
\begin{minipage}{0.33\hsize}
\resizebox{1.0\textwidth}{!}{
\includegraphics{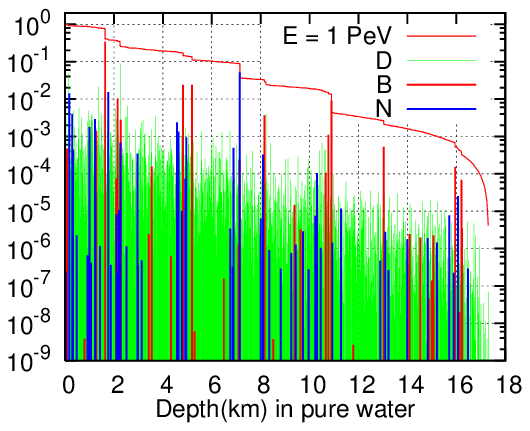}}	
\caption{The energy losses with the average-like range for $10^{15}$eV muons.}
\label{fig:ELA15}
\end{minipage}
\begin{minipage}{0.33\hsize}
\resizebox{1.0\textwidth}{!}{
\includegraphics{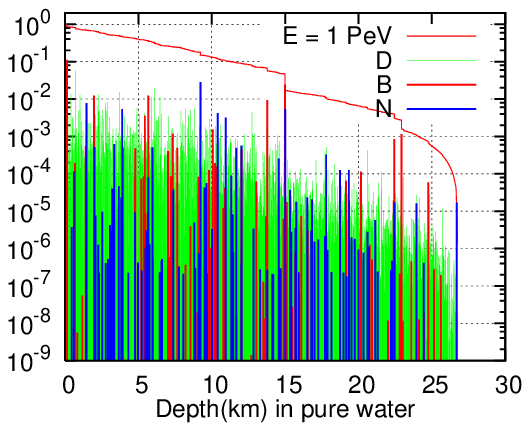}}	
\caption{The energy losses with the longest range for $10^{15}$eV muons.}
\label{fig:ELL15}
\end{minipage}
\end{tabular}
\end{center}
\end{figure}
\begin{figure}[h]
\begin{center}
\begin{tabular}{ccc}
\begin{minipage}{0.33\hsize}
\resizebox{1.0\textwidth}{!}{
\includegraphics{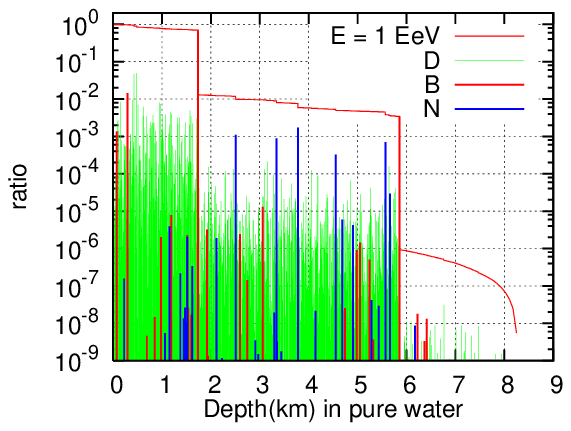}}	
\caption{The energy losses with the shortest range for $10^{18}$eV muons.}
\label{fig:ELS18}
\end{minipage}
\begin{minipage}{0.33\hsize}
\resizebox{1.0\textwidth}{!}{
\includegraphics{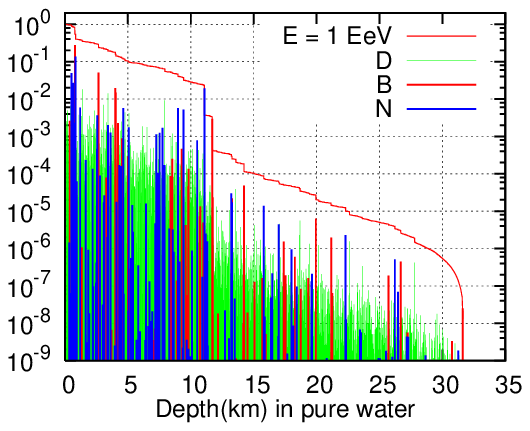}}	
\caption{The energy losses with the average-like range for $10^{18}$eV muons.}
\label{fig:ELA18}
\end{minipage}
\begin{minipage}{0.33\hsize}
\resizebox{1.0\textwidth}{!}{
\includegraphics{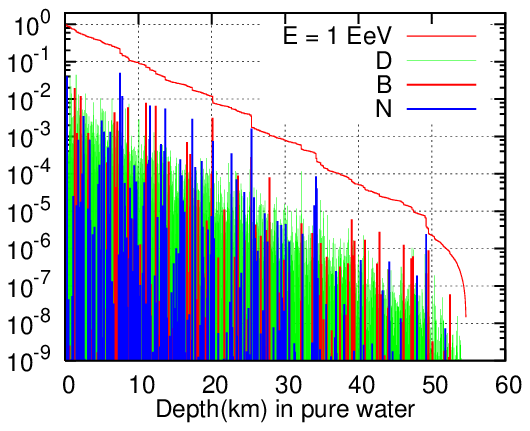}}	
\caption{The energy losses with the longest range for $10^{18}$ eV muons}
\label{fig:ELL18}
\end{minipage}
\end{tabular}
\end{center}
\end{figure}
Also, in Table 2 and Table 3, we summarize the characteristics of the events in Figure 17 to 25. Let us examine the characteristics of the events in Figures 17 to 25, combined with Tables 2 and 3. The energy losses for given primary energy of $10^{12}$eV due to different stochastic processes are given as the function of the depths traversed.

  In Figure 17, we give the case with the shortest range. The ordinate denotes the ratio of the energy loss to the primary energy and D,B,N denote the causes of energy losses, namely, direct electron pair production(D), bremsstrahlung(B) and photon nuclear interaction(N), respectively. The abscissa denotes the depths where the interactions concerned occur. The continuous lines (red lines) represent the muon energies at the corresponding depths.

  It is seen from Figure 17 and Tables 2 and 3 that this muon lose $\sim$10\% of the primary energy at $\sim$2 meters from the starting point and the almost of the primary energy by the bremsstrahlung  at $\sim$62 meters after several  negligible experiences of energy losses due to both the direct electron pair production and photo nuclear interaction. 99.9\% of the total energy loss is lost by only two bremsstrahlung. This is a good example of showing the catastrophic energy loss due to the bremsstrahlung.
\begin{table}[h]        
\begin{center}
\caption{The details of the characteristics on the muons with the shortest range, the average-like range ,the longest range and the average range.}
\label{tab:ELT}
\scalebox{0.6}[0.8]{
\begin{tabular}{c|c|c|c|c|c|c|c}
\hline\noalign{\smallskip}
	& Range & Energy loss & Number of   & Energy loss    & Number of   & Energy loss & Number of \\
$E_{p}=10^{12}eV$	&	[km]	& by brems  & interaction & by direct pair & interaction & by nuclear  & interaction \\
\noalign{\smallskip}\hline
$<$Average$>$& 2.39$\times 10^{0}$ & 1.09$\times 10^{11}$ & 4.67$\times 10^{0}$ & 1.70$\times 10^{11}$ & 2.39$\times 10^{2}$ & 4.43$\times 10^{10}$ & 3.37$\times 10^{0}$\\ \hline
Shortest    & 6.23$\times 10^{-2}$ &  9.81$\times 10^{11}$ & 2 & 4.73$\times 10^{8}$ &  4 & 3.99$\times 10^{8}$ & 1 \\ \hline
Average-like&  2.51$\times 10^{0}$ &  1.17$\times 10^{11}$ & 3 & 1.44$\times 10^{11}$ & 232 & 1.37$\times 10^{11}$ & 4 \\ \hline
Longest     &  9.29$\times 10^{0}$ &  6.09$\times 10^{10}$ & 8 & 1.44$\times 10^{11}$ & 288 & 1.66$\times 10^{9}$ & 2 \\ \hline
\noalign{\smallskip}
$E_{p}=10^{15}eV$\\ \noalign{\smallskip}\hline
$<$Average$>$& 1.72$\times 10^{1}$ & 3.38$\times 10^{14}$ & 4.64$\times 10^{1}$ & 4.96$\times 10^{14}$ & 6.56$\times 10^{3}$ & 1.61$\times 10^{14}$ & 5.31$\times 10^{1}$\\ \hline
Shortest    & 1.68$\times 10^{0}$ & 8.59$\times 10^{14}$ &  5 & 8.87$\times 10^{13}$ &   628 & 5.20$\times 10^{13}$ &  4\\ \hline
Average-like& 1.73$\times 10^{1}$ & 4.23$\times 10^{14}$ & 35 & 4.75$\times 10^{14}$ &  6363 & 9.67$\times 10^{13}$ & 56\\ \hline
Longest     & 3.16$\times 10^{1}$ & 1.82$\times 10^{14}$ & 70 & 7.50$\times 10^{14}$ & 10981 & 5.84$\times 10^{13}$ & 91\\ \hline
\noalign{\smallskip}
$E_{p}=10^{18}eV$\\ \noalign{\smallskip}\hline
$<$Average$>$&3.17$\times 10^{1}$&3.24$\times 10^{17}$&1.04$\times 10^{2}$&4.59$\times 10^{17}$&2.48$\times 10^{4}$&2.17$\times 10^{17}$&1.66$\times 10^{2}$\\ \hline
Shortest    & 8.27$\times 10^{0}$ & 7.04$\times 10^{17}$ &  33 & 2.91$\times 10^{17}$ &  7285 & 4.77$\times 10^{15}$ &  40\\ \hline
Average-like& 3.19$\times 10^{1}$ & 3.76$\times 10^{17}$ &  95 & 3.51$\times 10^{17}$ & 22395 & 1.64$\times 10^{17}$ & 173\\ \hline
Longest     & 5.48$\times 10^{1}$ & 7.52$\times 10^{16}$ & 190 & 7.61$\times 10^{17}$ & 45367 & 2.73$\times 10^{17}$ & 299\\ \hline
\noalign{\smallskip}
\end{tabular}
}
\end{center}
\end{table}

  On the other hand, in Figure 19, we show the case with the longest range. The range given by Figure 19 , $\sim$2700 meters, is far longer compared with $\sim$62 meters with the shortest range. It is clear from the figure and tables that 69.7\% of the total energy is lost by 288 direct electron pair production. Only 29.5\% of the total energy is lost 8 bremsstrahlung and the contribution from photo nuclear interaction is negligible.

  In Figure 18, we show the case with the average-like range. The definition of the average-like range denote the case whose range is the nearest to the average range which is obtained from the total number of the events (100,000 sampled events). This case shows that 36.2\% of the total energy is lost by 232 direct electron pair production, 34.4\% is lost by 4 photo nuclear interaction and 29.4\% is lost by 3 bremsstrahlung, while in the real average case, 52.6\% of the total energy is lost by 239 direct electron pair production, 33.7\% by bremsstrahlung and 13.7\% by photo nuclear interaction.

  In Figures 20 to 22, we show the similar relations for $10^{15}$ eV muons as shown in  Figures 17 to 19 and Tables 2 to 3. The shortest range, $\sim$1.6 kilometer (Figure 20), is far shorter compared with the longest one, $\sim$22 kilometers(Figure 22). It is seen from Figure 20 and Tables that bremsstrahlung plays a decisive role as the cause of catastrophic energy loss, too($\sim$80\% of total energy at $\sim$840 meters). 85.9\% of the total energy is lost by 5 bremsstrahlung, 8.87\% by 628 direct electron pair productions and 5.20\% by 4 photo nuclear interactions. In Figure 22, we give the case with the longest range. Here, large number of the direct electron pair production with rather small energy play an important role, similarly as shown in Figure 19. Here, 75.7.8\% of the total energy is lost by 10981 direct electron pair production, 18.4\% by 70 bremsstrahlung and 5.9\% by 91 photo nuclear interaction. In Figure 21, we give the case with the average like range.  Here, 47.8\% of the total energy is lost by 6363 direct electron pair productions, 42.5\% by 35 bremsstrahlung and 9.72\% by 56 photo nuclear interactions, while in the real averages, 49.8\% of the total energy loss is lost by 6560 direct electron pair productions, 34.0\% by 46.4 bremsstrahlung and 16.2\% by 53.1 photo nuclear interactions.
\begin{table}[h]        
\begin{center}
\caption{The Ratios of energies transferred from bremsstrahlung, direct electron pair production and photo nuclear interaction to the total energy loss.}
\label{tab:ratio}
\scalebox{1.0}[1.0]{
\begin{tabular}{c|c|c|c}
\hline 
$E_{p}=10^{12}eV$&Brems&Direct Pair&Nuclear\\ 
\hline
$<$Average$>$&3.37$\times 10^{-1}$&5.26$\times 10^{-1}$&1.37$\times 10^{-1}$\\ \hline
Shortest    &9.99$\times 10^{-1}$&4.81$\times 10^{-4}$&4.06$\times 10^{-4}$\\ \hline
Average-like&2.94$\times 10^{-1}$&3.62$\times 10^{-1}$&3.44$\times 10^{-1}$\\ \hline
Longest     &2.95$\times 10^{-1}$&6.97$\times 10^{-1}$&8.04$\times 10^{-3}$\\ 
\hline 
$E_{p}=10^{15}eV$\\ 
\hline
$<$Average$>$&3.40$\times 10^{-1}$&4.98$\times 10^{-1}$&1.62$\times 10^{-1}$\\ \hline
Shortest    &8.59$\times 10^{-1}$&8.87$\times 10^{-2}$&5.20$\times 10^{-2}$\\ \hline
Average-like&4.25$\times 10^{-1}$&4.78$\times 10^{-1}$&9.72$\times 10^{-2}$\\ \hline
Longest     &1.84$\times 10^{-1}$&7.57$\times 10^{-1}$&5.90$\times 10^{-2}$\\ \hline
$E_{p}=10^{18}eV$\\ 
\hline
$<$Average$>$&3.24$\times 10^{-1}$&4.59$\times 10^{-1}$&2.17$\times 10^{-1}$\\ \hline
Shortest    &7.04$\times 10^{-1}$&2.91$\times 10^{-1}$&4.77$\times 10^{-3}$\\ \hline
Average-like&4.22$\times 10^{-1}$&3.94$\times 10^{-1}$&1.84$\times 10^{-1}$\\ \hline
Longest     &6.78$\times 10^{-2}$&6.86$\times 10^{-1}$&2.46$\times 10^{-1}$\\ \hline
\end{tabular}
}
\end{center}
\end{table}
 
 In Figures 23 to 25,  we show the similar relations for $10^{18}$eV muons as shown in Figures 17 to 19. The case with shortest range in Figure 23 has strong contrast to that with the longest range. The manner of the energy loss in Figure 23 is drastic with two big catastrophic energy loss due to bremsstrahlung, while that in the Figure 25 is moderate with no catastrophic energy loss. The shortest range , $\sim$8 kilometers, is far shorter compared with the longest range, $\sim$54 kilometers. It is seen from Figure 23 and Tables that bremsstrahlung is a decisive role as the cause of catastrophic energy loss. 70.4\% of the total energy is lost by 33 bremsstrahlung, 29.1\% by direct electron pair productions and 0.477\% by 40 photo nuclear interactions. In Figure 25, we give the case with the longest range. Here, 68.8\% of the total energy is lost by 45367 direct electron pair productions, 24.6\% by 299 photo nuclear interactions and only 6.78\% by 190 bremsstrahlung in the complete absence of catastrophic energy losses. In Figure 24, we give the case average-like range. Here, 39.88\% of the total energy is lost by direct electron pair productions, 42.2 \% by 95 bremsstrahlung and 18.4\% by 173 photo nuclear interactions, while, in the real average's, 45.9\% of the total energy is lost by 24800 direct electron pair productions, 32.4\% by 104 bremsstrahlung and 21.2\% by 166 photo-nuclear interactions. Thus, it is concluded that the diversity among muon propagation with the same primary energy should be noticed.
%
\section{Cherenkov lights production due to both the energy losses by the muon(naked muons) and the accompanied cascade showers initiated by the muons concerned}
\label{sec:3}
 It should be noticed that the energy losses by high energy 
muons are never measured directly. Usually, in high energy 
neutrino astrophysics experiments in water(ice), they are measured via 
Cherenkov lights which are produced not only by the muon 
itself, but also accompanied by the cascade showers due to  bremsstrahlung, 
direct electron pair production and photo nuclear interaction, all 
of which are generated by the parent muons.
 When the muons traverse the matter, they lose their energies by 
bremsstrahlung, direct electron pair production and photo 
nuclear interaction in addition to ionization. These stochastic
processes produce cascade showers whose primary particle 
is a photon in bremsstrahlung and is an electron(a positron) in 
the direct electron pair production and photons decayed  from $\pi_{0}$ 
and others in photo nuclear interactions. These accompanied cascade showers 
produced by these stochastic processes are twisted around 
the traversing muons and these showers produce Cherenkov light.
 In this section, we discuss various quantities obtained from high energy 
muons, imaging the one-cubic kilometer detector for 
high energy neutrino astrophysics, something like Ice Cube 
in the Antarctic.

 We simulate exactly not only the both the interaction points and dissipated energies due to all stochastic processes, but also simulate exactly the accompanied cascade showers themselves due to these stochastic processes for the calculation of Cherenkov lights. Namely, the total Cherenkov lights due to both the muon itself and the accompanied cascade showers are exactly simulated. Here, we adopt the one-dimensional cascade showers under Approximation B \cite{Rossi} as cascade showers. Concretely speaking, we simulate cascade showers as exactly as possible so that the segments of the simulated electrons in the cascade showers are decided in both their locations and energies, which produce finally Cherenkov lights with the attenuation coefficient.
%
\subsection{The ratio of the Cherenkov lights production due to the accompanied cascade showers to the total Cherenkov lights}
\begin{figure}[t]
\begin{center}
\begin{tabular}{cc}
\begin{minipage}{0.45\hsize}
\resizebox{1.0\textwidth}{!}{
\includegraphics{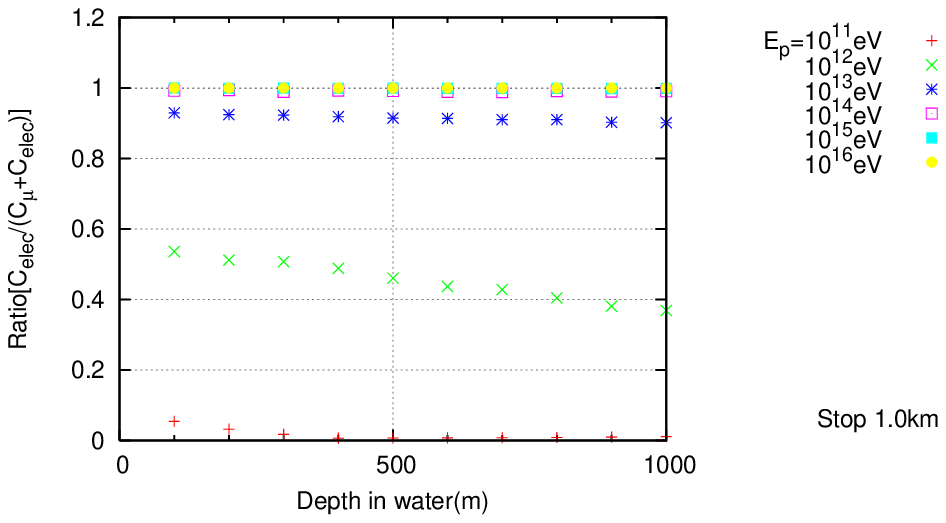}}	         
\caption{Ratio of Cherenkov lights due to the accompanied cascade showers to the total Cherenkov light}
\label{fig:RCR}
\end{minipage}
\begin{minipage}{0.4\hsize}
\resizebox{0.5\textwidth}{!}{
\includegraphics{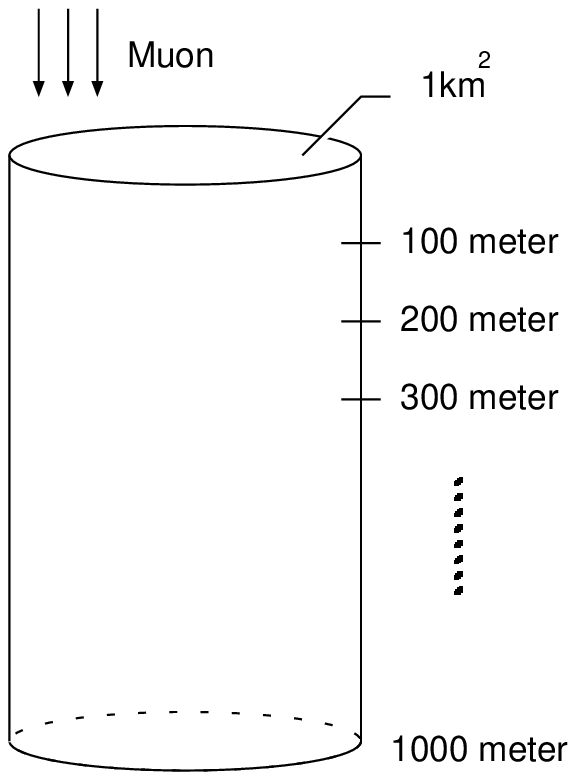}}	        
\caption{Observation points for the correlation between energy losses due to muon and the produced Cherenkov lights}
\label{fig:1kmcube}
\end{minipage}
\end{tabular}
\end{center}
\end{figure}

 In Figure \ref{fig:RCR}, we give the ratios of the Cherenkov lights due to the accompanied cascade showers to the Cherenkov lights due to (the accompanied cascade showers plus muon itself as the functions of the depth).
 Near $\sim10^{11}$eV,  most of  Cherenkov light  comes from muon itself(naked muon). Near $\sim10^{12}$eV,  about half of total Cherenkov light comes from muon itself muon.  Near $\sim10^{13}$eV,  90\% of the total Cherenkov light comes from the accompanied cascade showers. Above $10^{14}$eV, most of the total Cherenkov light is due to the accompanied cascade showers.
%
\subsection{The correlation between the total Cherenkov lights and the corresponding energy losses}
\begin{figure}[h]
\begin{tabular}{cc}
\begin{minipage}{0.33\hsize}
\resizebox{1.0\textwidth}{!}{
\includegraphics{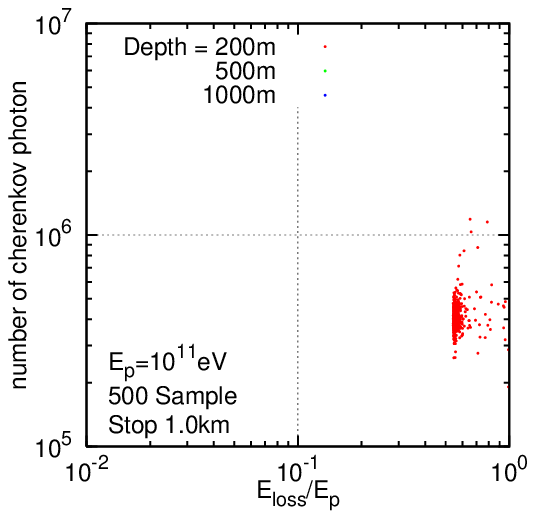}}	       
\caption{Correlation diagram for $10^{11}$eV muons}
\label{fig:CEL11}
\end{minipage}
\begin{minipage}{0.33\hsize}
\resizebox{1.0\textwidth}{!}{
\includegraphics{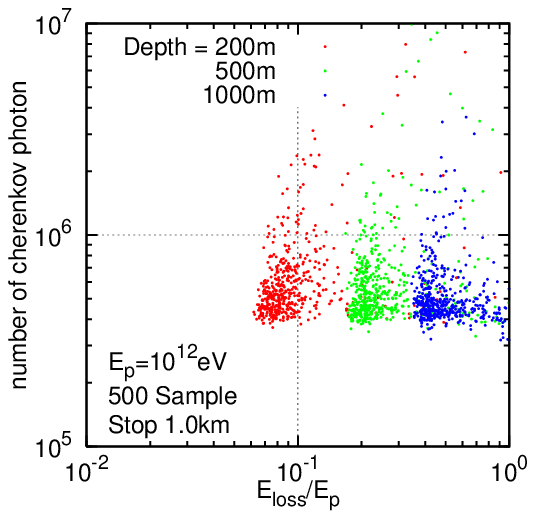}}	       
\caption{Correlation diagram for $10^{12}$eV muons}
\label{fig:CEL12}
\end{minipage}
\end{tabular}
\end{figure}

\begin{figure}[h]
\begin{tabular}{ccc}
\begin{minipage}{0.33\hsize}
\resizebox{1.0\textwidth}{!}{
\includegraphics{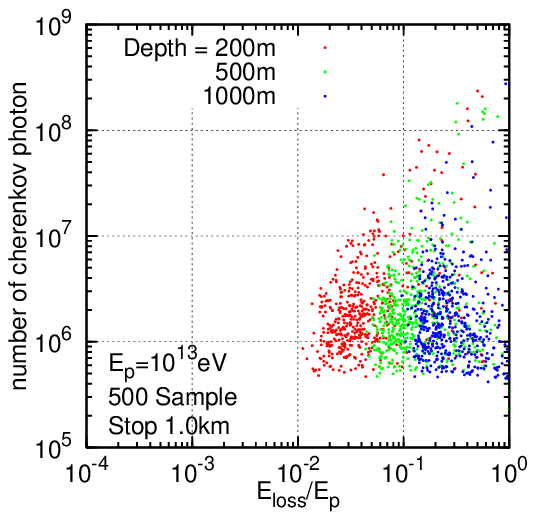}}	       
\caption{Correlation diagram for $10^{13}$eV muons}
\label{fig:CEL13}
\end{minipage}
\begin{minipage}{0.33\hsize}
\resizebox{1.0\textwidth}{!}{
\includegraphics{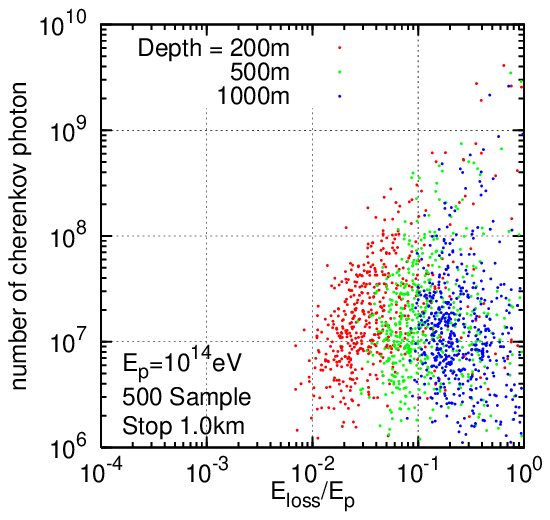}}	       
\caption{Correlation diagram for $10^{14}$eV muons}
\label{fig:CEL14}
\end{minipage}
\begin{minipage}{0.33\hsize}
\resizebox{1.0\textwidth}{!}{
\includegraphics{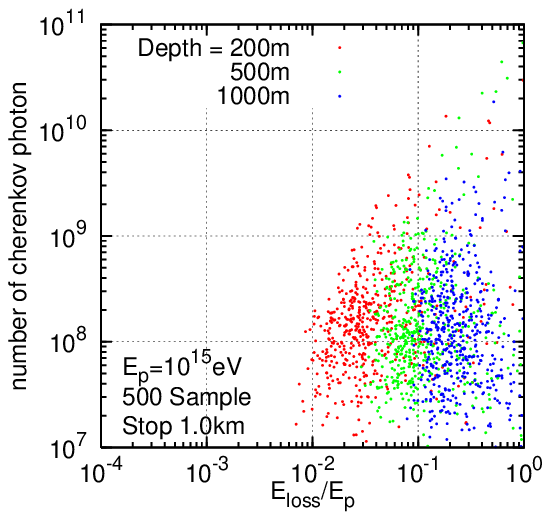}}	       
\caption{Correlation diagram for $10^{15}$eV muons}
\label{fig:CEL15}
\end{minipage}
\end{tabular}
\end{figure}

\begin{figure}[h]
\begin{tabular}{ccc}
\begin{minipage}{0.33\hsize}
\resizebox{1.0\textwidth}{!}{
\includegraphics{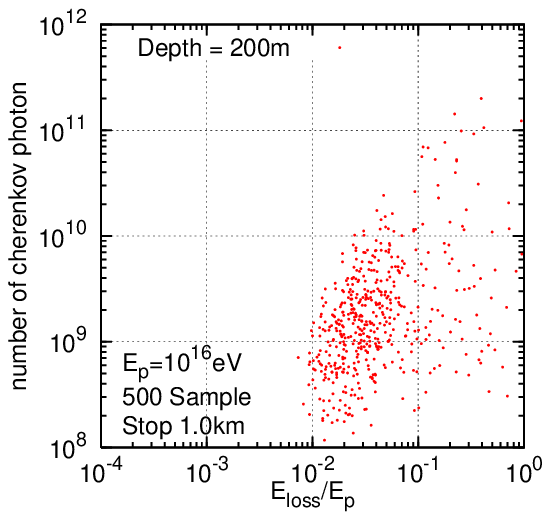}}	  
\caption{Correlation diagram at 200 meters for $10^{16}$eV muons}
\label{fig:CEL16_200m}
\end{minipage}
\begin{minipage}{0.33\hsize}
\resizebox{1.0\textwidth}{!}{
\includegraphics{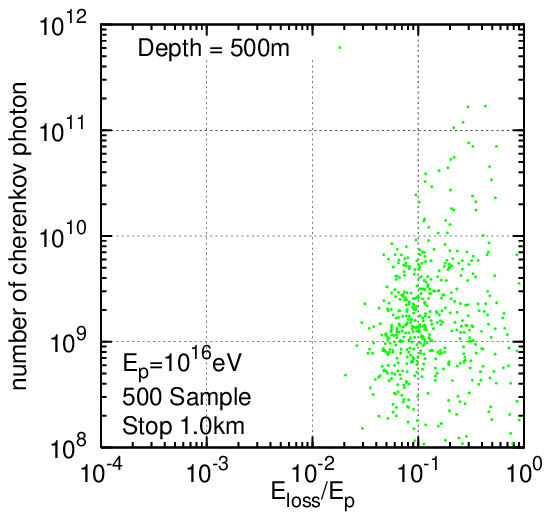}}	  
\caption{Correlation diagram at 500 meters for $10^{16}$eV muons}
\label{fig:CEL16_500m}
\end{minipage}
\begin{minipage}{0.33\hsize}
\resizebox{1.0\textwidth}{!}{
\includegraphics{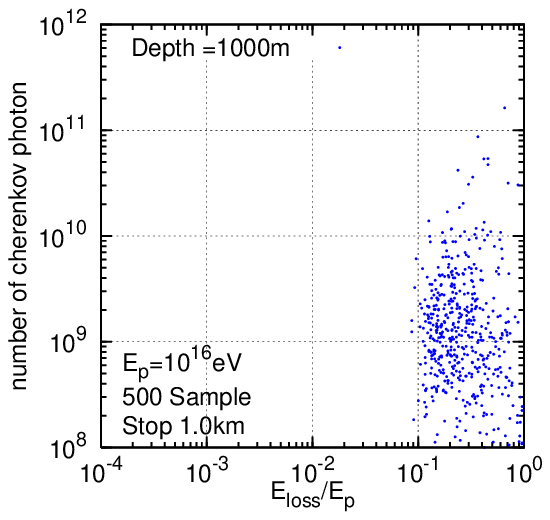}}	     
\caption{Correlation diagram at 1000 meters for $10^{16}$eV muons}
\label{fig:CEL16_1000m}
\end{minipage}
\end{tabular}
\end{figure}

 We examine the following correlations at observation points as shown in Figure \ref{fig:1kmcube}. 
In Figures \ref{fig:CEL11} to \ref{fig:CEL16_1000m}, we give the correlation diagram at observation points, such as, 200, 500 and 1000 meters from the incident points in the case of $10^{11}$ to $10^{16}$eV muons, respectively, between the energy losses due to the stochastic processes in addition to ionization and Cherenkov lights which are produced by both the muon themselves and their accompanied cascade showers. The attenuation coefficient is considered in the propagation Cherenkov lights. 
 Here, [ energy loss ] denotes that the energy dissipated by the muon while traversing through some distance (for example, 200 meters) due to bremsstrahlung, direct electron pair production and photo nuclear interaction  in addition to ionization loss. [Cherenkov lights] denote the measured Cherenkov lights at some depth (for example, 200 meters) which are produced by not only the muon itself, but also, the accompanied cascade showers due to the possible stochastic processes taking into account of the attennation effect of Cherenkov lights. 
 
 In Figure \ref{fig:CEL11}, we give the correlation between the energy loss and Cherenkov lights up to 200 meters from the starting point initiated by $10^{11}$eV muon(naked muon). As seen from Figure \ref{fig:RCR}, the most Cherenkov lights are produced by muon itself and the smaller part is produced by the accompanied cascade shower. Red points denote the correlation at 200 meters It is understood from the figure that more dense region may come from the muon itself and more weaker dense region may come from accompanied cascade showers. We cannot observe Cherenkov lights at both 500 meters and 1000 meters, because the energy of $10^{11}$eV is too small to detect at 500 meters and 1000 meters. 
 
 In Figures \ref{fig:CEL12} to \ref{fig:CEL16_1000m}, green points and blue pints stand for the correlation at 500 meters and 1000 meters, respectively. In Figure \ref{fig:CEL12}, we give the similar diagram as shown in Figure \ref{fig:CEL11} and ,there, we give the correlations at 200 meters, 500 meters and 1000 meters. As seen from Figure \ref{fig:RCR}, $\sim$ half of the total Cherenkov lights may be due to muon itself 's origin and $\sim$the other half may be accompanied cascade showers' origin. It is clear from the figure that the domain for the correlation at 200 meters is larger than those at 500 meters and 1000 meters, the meaning of which shows bigger fluctuation at 200 meters compared with observation points.  
 
 The larger part of the energy losses may be initiated by the accompanied cascade showers. At $10^{13}$eV, as seen from Figure \ref{fig:RCR}, $\sim$ 90 \% of the total Cherenkov lights may be produced by the accompanied cascade showers and the domain for the correlation at different depths  begin to  overlap. Above $10^{14}$eV, almost of  Cherenkov lights are produced by the accompanied cascade showers. As the developments of the cascade showers are suffered from the fluctuation. It is clear from Figures \ref{fig:CEL14} and \ref{fig:CEL15} that the boundaries  of the domains for the correlation at $10^{14}$eV and $10^{15}$eV become obscure. 
 
 In Figures \ref{fig:CEL16_200m} to \ref{fig:CEL16_1000m}, we give the correlation at different depths, say 200 meters, 500 meters and 1000 meters for $10^{16}$eV muon, separately and respectively. The comparison among Figures shows that fluctuation in the energy losses become decrease, as the depths increase. Also, it should be noticed from the figures that the degree of the fluctuation in the total Cherenkov lights is bigger than that of the energy losses. 
 
 The fluctuation of the total Cherenkov lights is directly related to that of the accompanied cascade showers which start different depth having at different primary energies.
%
\subsection{Fluctuation in the energy loss distribution for given primary energies at the depths, 200 meters, 500 meters and 1000 meters.}
\begin{figure}[h]
\begin{tabular}{ccc}
\begin{minipage}{0.33\hsize}
\resizebox{1.0\textwidth}{!}{
\includegraphics{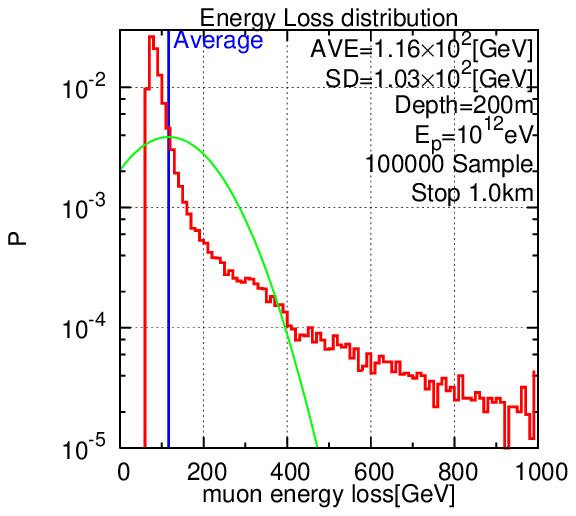}}			
\caption{Energy loss distribution at 200 meters for $10^{12}$eV muons}
\label{fig:ELD12_200mlog}
\end{minipage}
\begin{minipage}{0.33\hsize}
\resizebox{1.0\textwidth}{!}{
\includegraphics{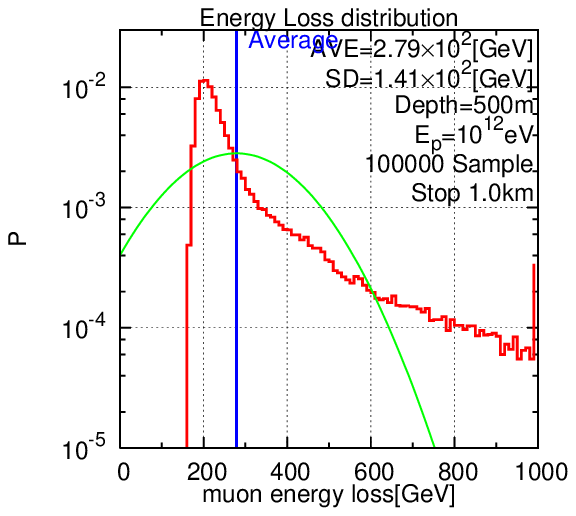}}		
\caption{Energy loss distribution at 500 meters for $10^{12}$eV muons}
\label{fig:ELD12_500mlog}	
\end{minipage}
\begin{minipage}{0.33\hsize}
\resizebox{1.0\textwidth}{!}{
\includegraphics{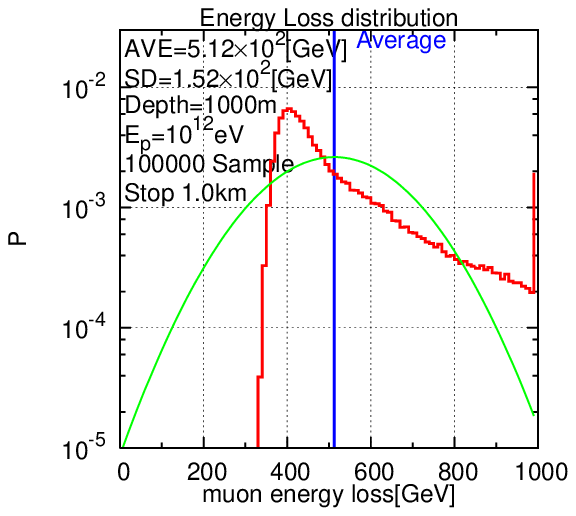}}	
\caption{Energy loss distribution at 1000 meters for $10^{12}$eV muons}
\label{fig:ELD12_1000mlog}
\end{minipage}
\end{tabular}
\end{figure}

\begin{figure}[h]
\begin{tabular}{ccc}
\begin{minipage}{0.33\hsize}
\resizebox{1.0\textwidth}{!}{
\includegraphics{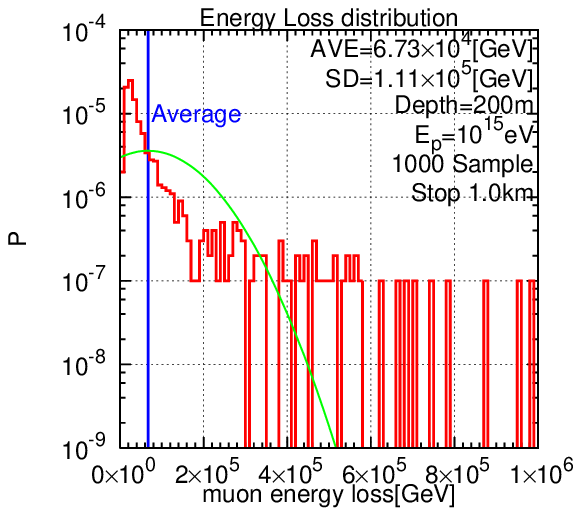}}		
\caption{Energy loss distribution at 200 meters for $10^{15}$eV muons}
\label{fig:ELD15_200mlog}
\end{minipage}
\begin{minipage}{0.33\hsize}
\resizebox{1.0\textwidth}{!}{
\includegraphics{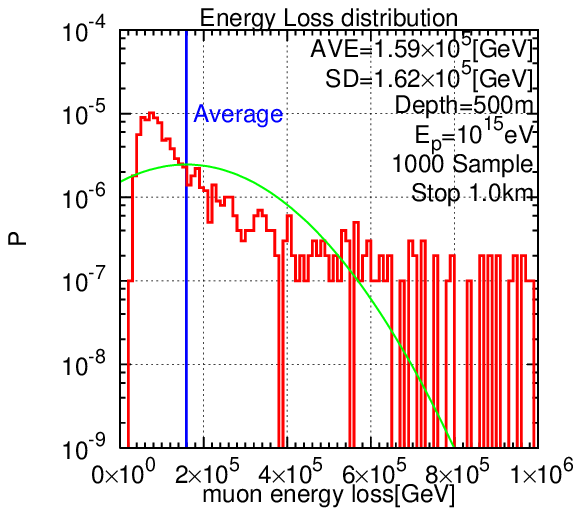}}		
\caption{Energy loss distribution at 500 meters for $10^{15}$eV muons.}
\label{fig:ELD15_500mlog}
\end{minipage}
\begin{minipage}{0.33\hsize}
\resizebox{1.0\textwidth}{!}{
\includegraphics{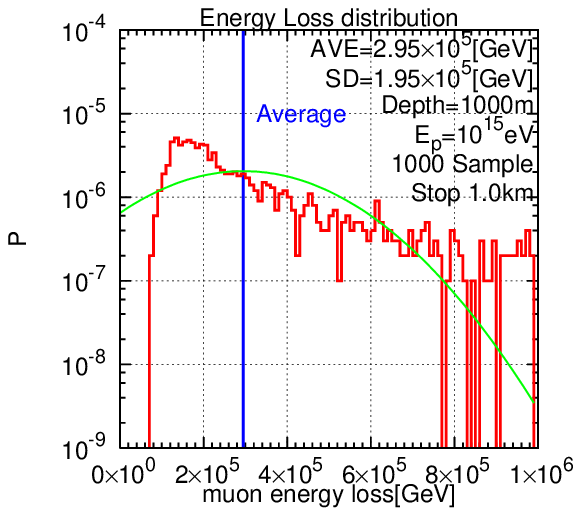}}	
\caption{Energy loss distribution at 1000 meters for $10^{15}$eV muons}
\label{fig:ELD15_1000mlog}
\end{minipage}
\end{tabular}
\end{figure}

\begin{figure}[h]
\begin{tabular}{ccc}
\begin{minipage}{0.33\hsize}
\resizebox{1.0\textwidth}{!}{
\includegraphics{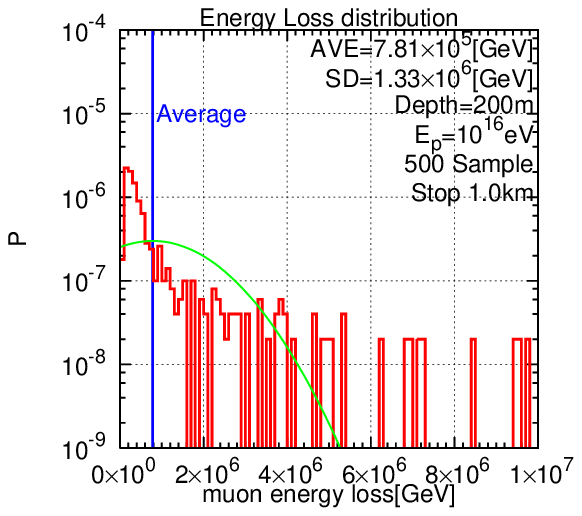}}		
\caption{Energy loss distribution at 200 meters for $10^{16}$eV muons}
\label{fig:ELD16_200mlog}
\end{minipage}
\begin{minipage}{0.33\hsize}
\resizebox{1.0\textwidth}{!}{
\includegraphics{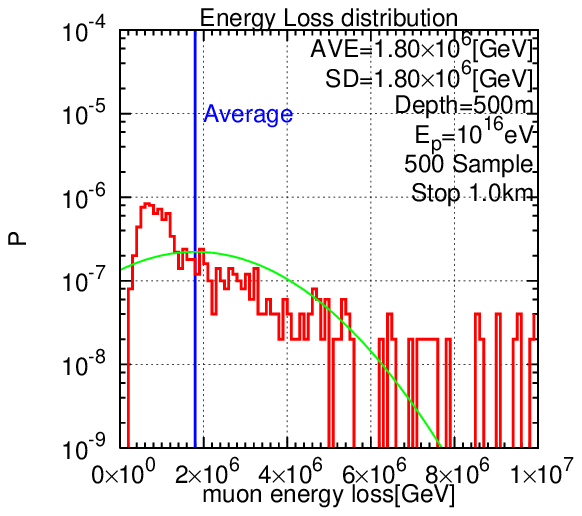}}		
\caption{Energy loss distribution at 500 meters for $10^{16}$eV muons}
\label{fig:ELD16_500mlog}
\end{minipage}
\begin{minipage}{0.33\hsize}
\resizebox{1.0\textwidth}{!}{
\includegraphics{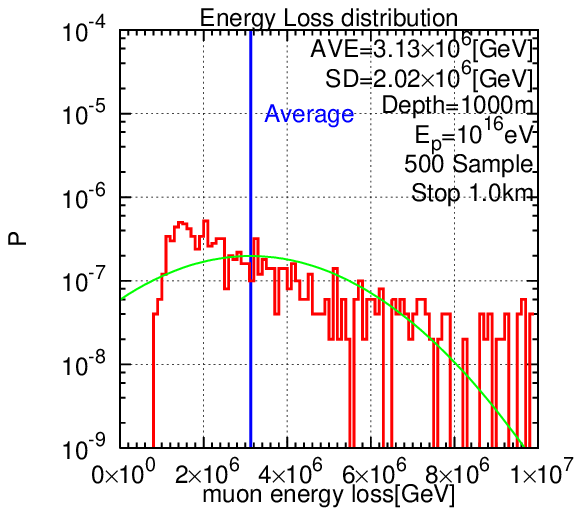}}		
\caption{Energy loss distribution at 1000 meters for $10^{16}$eV muons}
\label{fig:ELD16_1000mlog}
\end{minipage}
\end{tabular}
\end{figure}

 In Figures \ref{fig:ELD12_200mlog} to \ref{fig:ELD12_1000mlog}, we give the energy loss distribution for $10^{12}$eV at 200 meters, 500 meters and 1000 meters, respectively. In Figures \ref{fig:ELD15_200mlog} to \ref{fig:ELD15_1000mlog}, and Figures \ref{fig:ELD16_200mlog} to \ref{fig:ELD16_1000mlog}, we give the similar distributions for $10^{15}$eV muons and for $10^{16}$eV, respectively. In these Figures, we add the normal distributions with the same average values and same standard deviations which are obtained from the real distributions in Figures \ref{fig:ELD12_200mlog} to \ref{fig:ELD16_1000mlog}. It is clear from these figures that the normal distributions for the energy losses never express the real distributions on the contrast to the cases of range fluctuation (see, Eq.(\ref{PRE})). The cause of the fluctuation in the energy losses comes from the compound effect of the both fluctuation in the interaction points due to different stochastic processes and the fluctuation of the energy release from different stochastic processes.
\subsection{Fluctuation in the total Cherenkov lights quantities for given primary primaries energy at the depths, 200 meters, 500 meters and 1000 meters.}
\begin{figure}[b]
\begin{tabular}{ccc}
\begin{minipage}{0.33\hsize}
\resizebox{1.0\textwidth}{!}{
\includegraphics{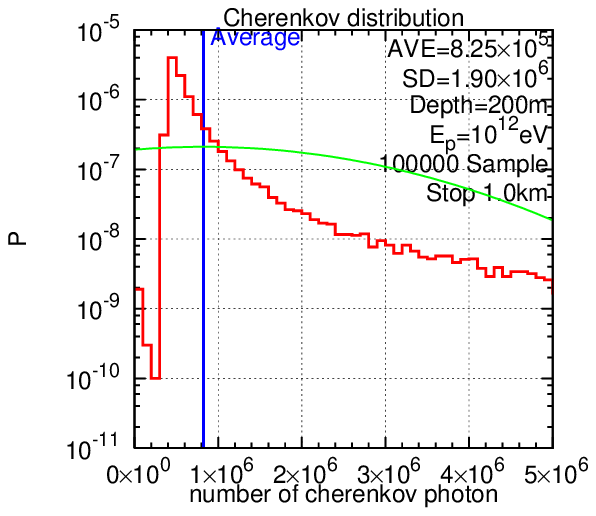}}	
\caption{Cherenkov lights distribution at 200 meters for $10^{12}$eV muons}
\label{fig:CHD12_200mlog}
\end{minipage}
\begin{minipage}{0.33\hsize}
\resizebox{1.0\textwidth}{!}{
\includegraphics{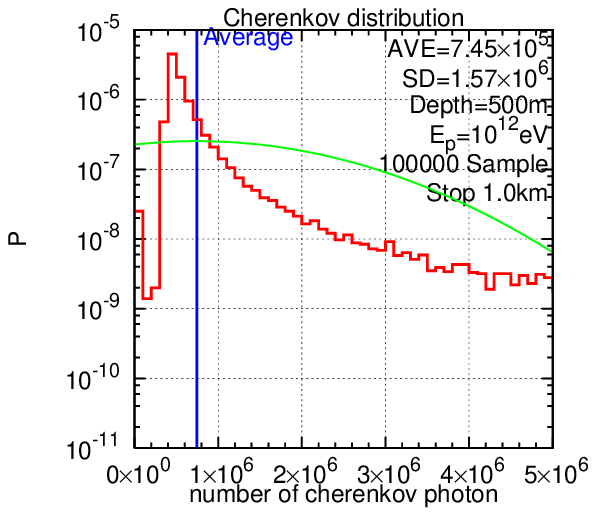}}	
\caption{Cherenkov lights distribution at 500 meters for $10^{12}$eV muons}
\label{fig:CHD12_500mlog}
\end{minipage}
\begin{minipage}{0.33\hsize}
\resizebox{1.0\textwidth}{!}{
\includegraphics{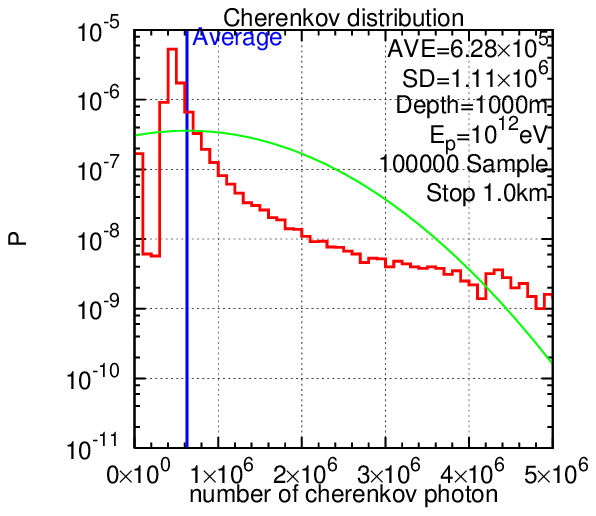}}	
\caption{Cherenkov lights distribution at 1000 meters for $10^{12}$eV muons}
\label{fig:CHD12_1000mlog}
\end{minipage}
\end{tabular}
\end{figure}

\begin{figure}[h]
\begin{tabular}{ccc}
\begin{minipage}{0.33\hsize}
\resizebox{1.0\textwidth}{!}{
\includegraphics{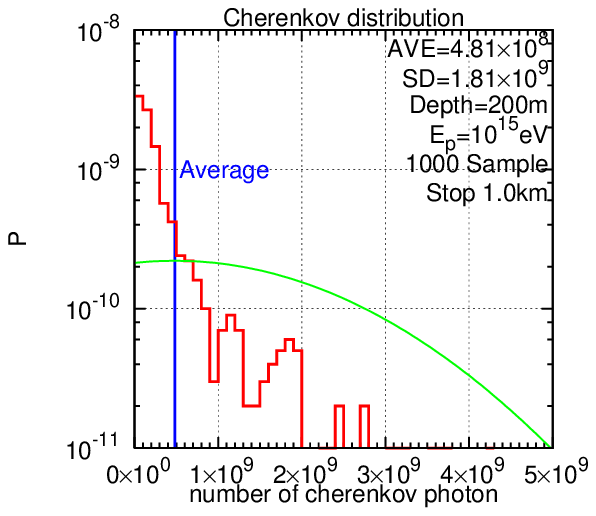}}	
\caption{Cherenkov lights distribution at 200 meters for $10^{15}$eV muons}
\label{fig:CHD15_200mlog}
\end{minipage}
\begin{minipage}{0.33\hsize}
\resizebox{1.0\textwidth}{!}{
\includegraphics{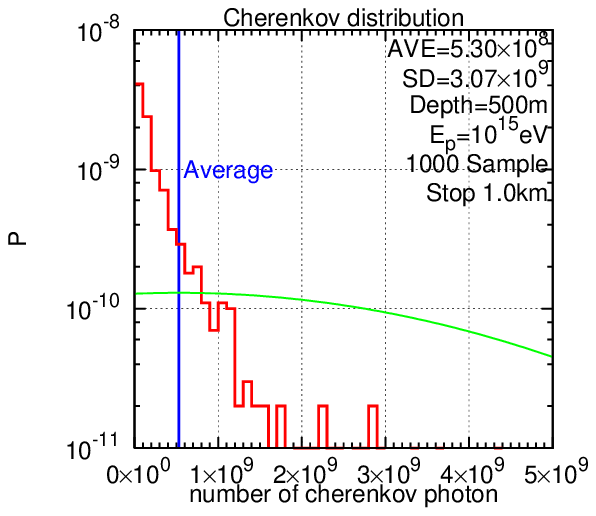}}	
\caption{Cherenkov lights distribution at 500 meters for $10^{15}$eV muons}
\label{fig:CHD15_500mlog}
\end{minipage}
\begin{minipage}{0.33\hsize}
\resizebox{1.0\textwidth}{!}{
\includegraphics{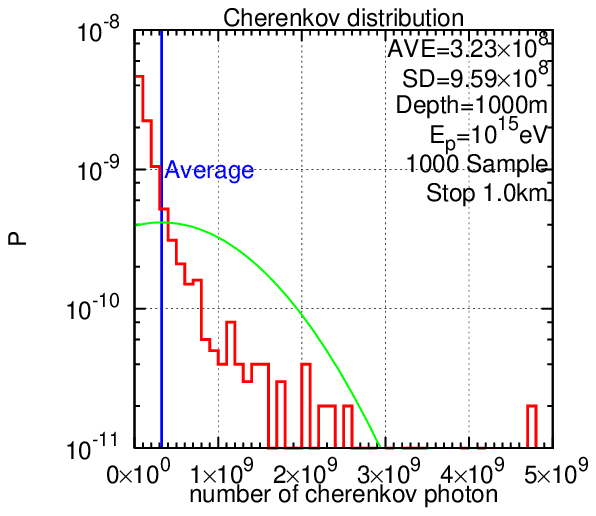}}	
\caption{Cherenkov lights distribution at 1000 meters for $10^{15}$eV muons}
\label{fig:CHD15_1000mlog}
\end{minipage}
\end{tabular}
\end{figure}

\begin{figure}[h]
\begin{tabular}{ccc}
\begin{minipage}{0.33\hsize}
\resizebox{1.0\textwidth}{!}{
\includegraphics{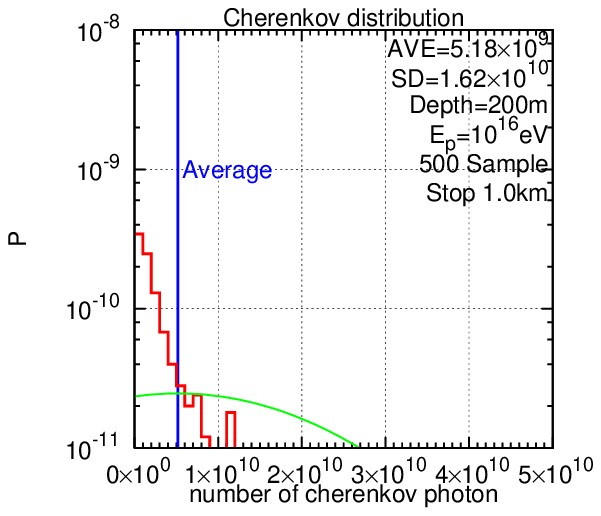}}		
\caption{Cherenkov lights distribution at 200 meters for $10^{16}$eV muons}
\label{fig:CHD16_200mlog}
\end{minipage}
\begin{minipage}{0.33\hsize}
\resizebox{1.0\textwidth}{!}{
\includegraphics{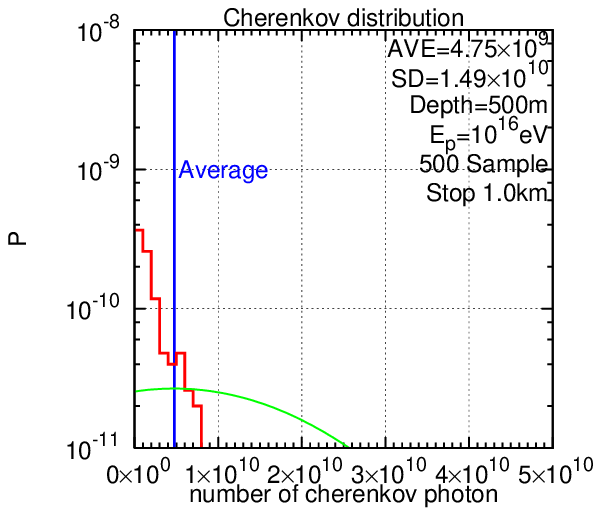}}		
\caption{Cherenkov lights distribution at 500 meters for $10^{16}$eV muons}
\label{fig:CHD16_500mlog}
\end{minipage}
\begin{minipage}{0.33\hsize}
\resizebox{1.0\textwidth}{!}{
\includegraphics{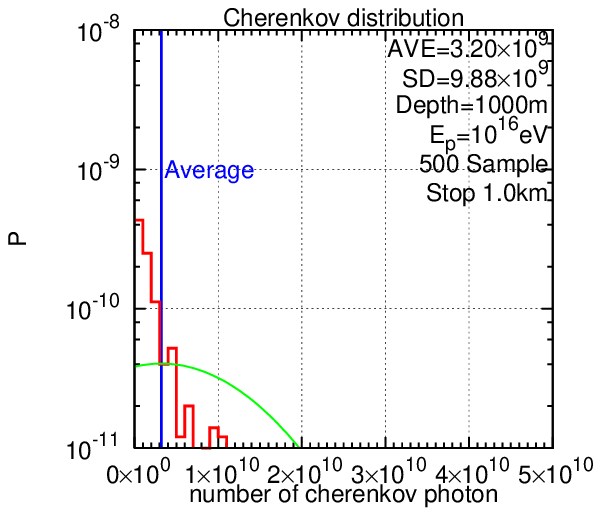}}		
\caption{Cherenkov lights distribution at 1000 meters for $10^{16}$ eV muons}
\label{fig:CHD16_1000mlog}
\end{minipage}
\end{tabular}
\end{figure}

 The primary muons are dissipated their energy by bremsstrahlung, direct electron pair production and photo nuclear interaction in addition to the ionization loss. Thus, these stochastic processes are the origin of the accompanied cascade showers which finally produce Cherenkov lights in addition to the Cherenkov lights due to the muon itself(naked muon).
  It is clear from the Figures \ref{fig:CHD12_200mlog} to \ref{fig:CHD16_1000mlog} that the Cherenkov lights quantities thus obtained by stochastic processes are widely distributed due to the complicated compound fluctuation effect coming from various stochastic sources. The normal distribution whose average values and the standard deviation are taken from the real distributions are given with the real distributions for the comparison in Figures \ref{fig:CHD12_200mlog} to \ref{fig:CHD16_1000mlog}. It is easily understood that such normal distributions never reflect upon the real situations. The correlation between the energy losses and Cherenkov lights are obtained from the combination of the energy losses as shown in Figures \ref{fig:ELD12_200mlog} to \ref{fig:ELD16_1000mlog} with the corresponding Figures \ref{fig:CHD12_200mlog} to \ref{fig:CHD16_1000mlog}.
\\
\subsection{Transition curves of the averaged Cherenkov lights for different primary energies}
In figure 54, we give the average transition curve for Cherenkov lights for $10^{12}$eV muon  in the KM3 detector. The sampled number of the events is 500. The uncertainty bars in the figures show the range within which 68 \% of the total number is included which correspond to one $\sigma$ in the normal distribution. As shown in Figures 45 to 56, Cherenkov lights distributions are far deviated from the corresponding normal distribution and ,therefore, the ranges for the uncertainty too large those of the normal distributions. Also, gradual decreases of the slopes in the average transition curves denote that the energy losses up to the depths cannot be neglected compared with the primary energies over 1 kilometers.
 In Figures 55 and 57, we give the similar transition curves to Figure 54 for $10^{13}$eV and $10^{15}$eV, respectively. Compared these curves with the curve in Figure 54, it is seen that the average energy losses are almost independent on the behaviors of the muons concerned over 1 kilometers. However, the uncertainties around the average values become large compared with that in Figure 55, which denotes bigger uncertainty as the increase of the primary energies. It should be noticed from these figures that there are the uncertainties by one order of magnitude above $10^{13}$eV muons.
\begin{figure}[h]
\begin{center}
\begin{tabular}{cc}
\begin{minipage}{0.5\hsize}
\resizebox{0.9\textwidth}{!}{
\includegraphics{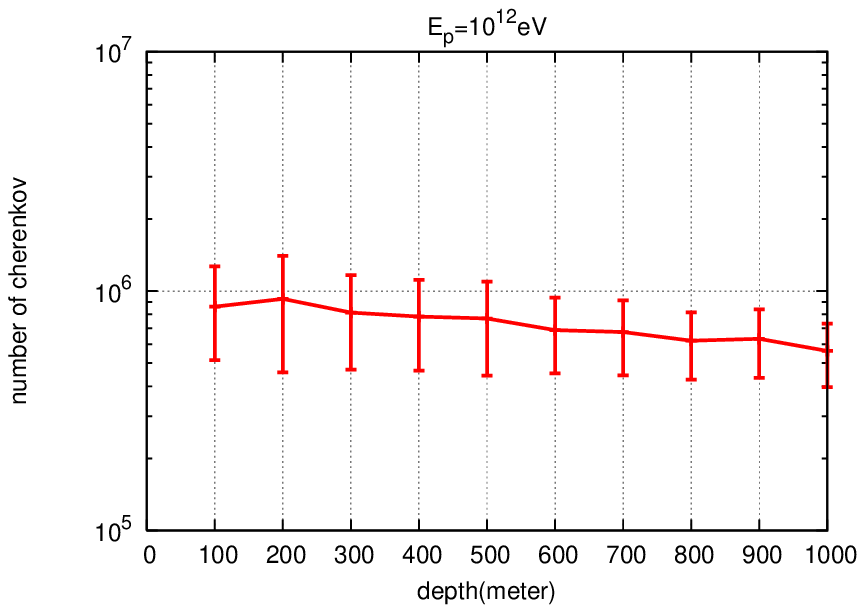}}		
\caption{Transition Curve of  the Averaged Cherenkov lights and their one sigma with the primary energy of $10^{12}$ eV}
\label{fig:Ep10_12_68}
\end{minipage}
\begin{minipage}{0.5\hsize}
\resizebox{0.9\textwidth}{!}{
\includegraphics{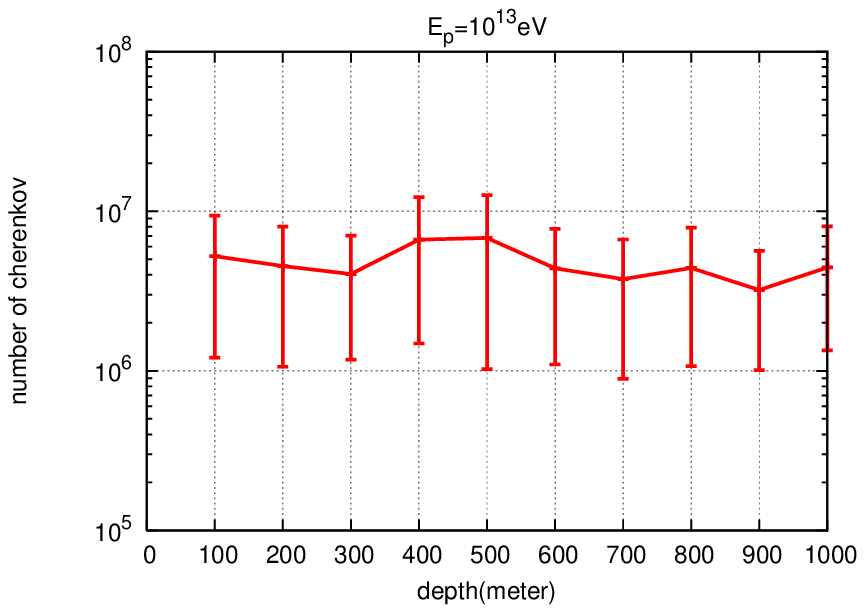}}	
\caption{Transition Curve of  the Averaged Cherenkov lights and their one sigma with the primary energy of $10^{13}$ eV}
\label{fig:Ep10_13_68}
\end{minipage}
\end{tabular}
\end{center}
\end{figure}
\begin{figure}[h]
\begin{center}
\begin{tabular}{cc}
\begin{minipage}{0.5\hsize}
\resizebox{0.9\textwidth}{!}{
\includegraphics{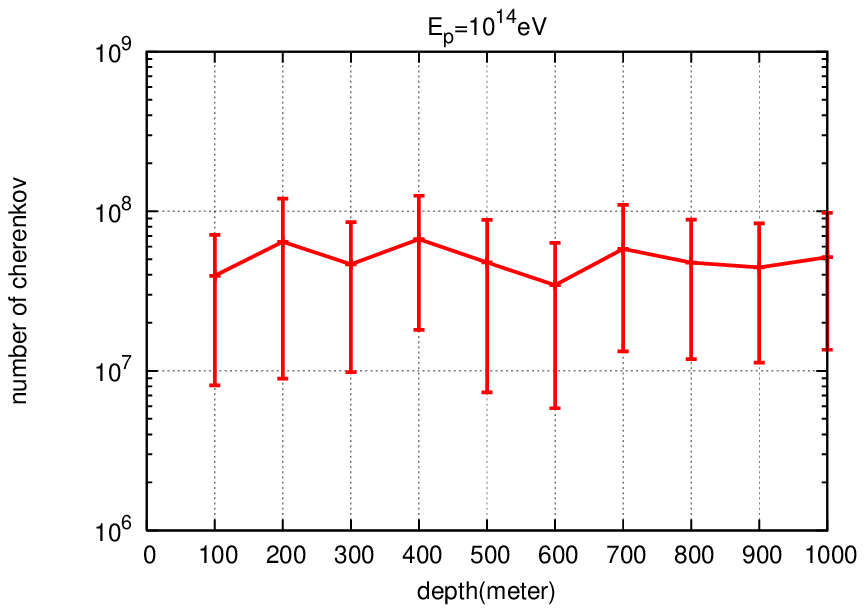}}		
\caption{Transition curve of the averaged Cherenkov lights and their one sigma with the primary energy of $10^{14}$ eV}
\label{fig:Ep10_14_68}
\end{minipage}
\begin{minipage}{0.5\hsize}
\resizebox{0.9\textwidth}{!}{
\includegraphics{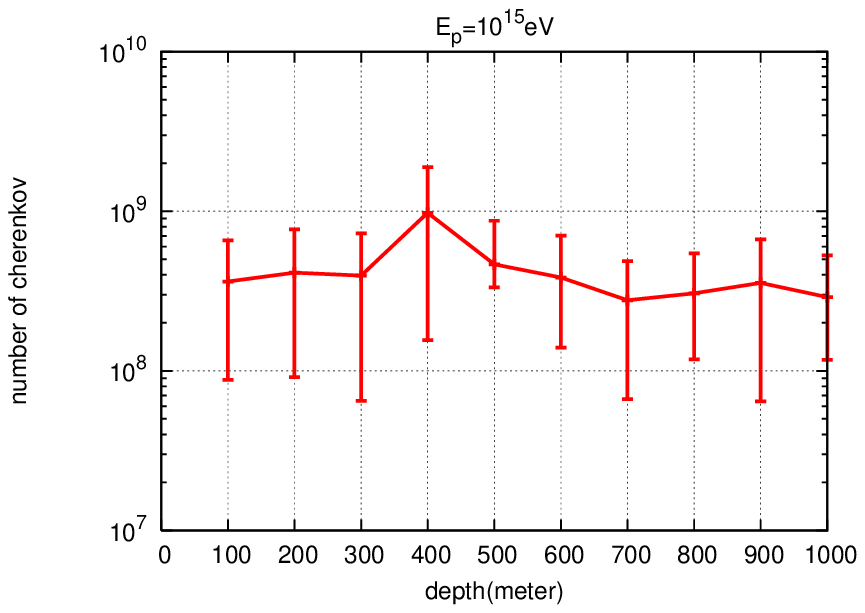}}	
\caption{Transition curve of the averaged Cherenkov lights and their one sigma with the primary energy of $10^{15}$ eV}
\label{fig:Ep10_15_68}
\end{minipage}
\end{tabular}
\end{center}
\end{figure}

\section{Discussions and Summary}
\label{sec:4}
 The validity of our Monte Carlo simulation procedure (\textit{the integral method}) has been proven to be right by two different methods. One method is analytical method (see, the Appendix) and the other is different Monte Carl procedure (\textit{the differential method}).(see, Figures 4 to 6).\\
 We apply our method to get the survival probabilities for different energies ($10^{12}$eV, $10^{15}$eV and $10^{18}$eV). Also, we have obtained the range distributions for different primary energies ($10^{11}$eV to $10^{18}$eV) and we have shown they can be expressed as the normal distribution well (see, Table 1).
 We have given the three typical muon behaviors [with the shortest range], [with the average-like range] and [with the longest range] for the different primary energies ($10^{12}$eV, $10^{15}$eV and $10^{18}$eV) to clarify the diversity among the muons' ranges. (Figure 17 to 25) and have summarized their characteristics in Table 2 and Table 3.\\
 In addition to the fluctuation of high energy muons, we have investigated fluctuations the total Cherenkov lights produced by muons. The total Cherenkov lights consists of two parts, namely, The Cherenkov lights due to the naked muons themselves and the Cherenkov lights through the accompanied cascade showers due to bremsstrahlung, direct electron pair production and photo nuclear interaction due to the muons. Above $10^{14}$eV, almost of the Cherenkov lights are produced exclusively by the accompanied cascade showers(See, Figure 26). The fluctuation of the Cherenkov lights due to the accompanied cascade showers come from not only the fluctuations of their interaction points due to the different stochastic processes, but also fluctuations of the transferred energies due to these processes due to the muons and the fluctuation on the development of the accompanied cascade showers.
We have simulated all possible stochastic processes related to the production of the Cherenkov lights as exactly possible and have obtained the dependence of the average Cherenkov lights on the depths traversed (the average transition curves for the Cherenkov lights )(Figure 54 to 57). It is clear from the Figures that fluctuation of the Cherenkov lights at each depth are very big, the ranges for uncertainties which are of one order of the magnitude. From these transition curves, we can estimate primary energies within some allowance, which is supposed to be one factor 5 order, really big uncertainty. 
 Here, it is adequate for us to mention the essential difference between \textit{the differential method}[5],[6],[7],[8],[9] and \textit{the integral method}[12],[22] as for the Cherenkov lights propagations. 
 Here, we return to the fundamental equation for high energy muon in order to focus on the dissipated energies by the muons. Again, the equation is given as,
\begin{eqnarray}  
 \left[\frac{dE}{dx}\right]_{rad}&=&\left[\frac{dE}{dx}\right]_{soft}+\left[\frac{dE}{dx}\right]_{rad} \nonumber \\
 &=&\frac{N}{A}E\int_{0}^{v_{cut}}dv\cdot v\frac{\sigma\left(v,E\right)}{dv}+\frac{N}{A}E\int_{v_{cut}}^{1}dv\cdot v\frac{d\sigma\left(v,E\right)}{dv}
\end{eqnarray}

In \textit{the differential method}, the Monte Carlo procedure is made on the hard part of Eq.(8) including the integral from $v_{cut}$ to $1$. The specified value of $v_{cut}$ is not so sensitive for the conclusions on the muon propagations, although usually $v_{cut}$=$10^{-2}$ to $10^{-4}$. Because, the total amount of the energy loss by the muon concerned is the same, irrespective of $v_{cut}$ in the integrals of Eq.(7). Therefore, the average behaviors of the high energy muons obtained by \textit{the differential method} are same as obtained by \textit{the integral method}(see, Figures 4 to 6). However, the shape of the energy loss spectrum obtained by \textit{the differential method} are expected to be different from those obtained by \textit{the integral method}, because the introduction of constant $v_{cut}$ for different primary muon energies into \textit{the differential method} may make the real energy loss spectrum deform 
\footnote{Exactly speaking, in \textit{the integral method}, $v_{cut}$ is introduced in the bremsstrahlung to avoid the catastrophic energy loss. However, $v_{cut}$ is taken as $v_{cut} \ll E_{min}/E_{0}$, where $E_{0}$ and $E_{min}$ denote the primary energy and the minimum energy for observation (1 Gev). Usually, we adopt $10^{-4}$ to $10^{-10}$ as $v_{cut}$. Consequently, we can reproduce the differential cross sections for bremsstrahlung exactly above $E_{min}$.
}
.

The energy loss spectrum at different depths are constructed through the compound processes in which the fluctuations of the interaction points due to the different stochastic processes and fluctuations in the transferred energies due to the different stochastic processes are exactly considered. The energy loss spectrum thus constructed are the origins of Cherenkov lights spectrum. Detailed speaking, the Cherenkov lights measured at some depth are the product by the complicated aggregate of the cascade showers whose starting points and whose primary energies are different due to the different stochastic processes. The muon energy loss spectrum should be considered in such the framework.

As for the muon energy spectrum, we have another matter to be considered related to \textit{the differential method}. In \textit{the differential method}, one introduces the constant value of $v_{cut}$ for the Monte carlo procedure. Here, it should be noticed that the primary muons have energy spectrum.   Suppose that when $E$ is greater than $E\times v_{cut}$, $E$ may lie in the radiation part of Eq.(1) in \textit{the differential method}. However, the $E$ with the same energy may lie the soft ( continuous ) part, if their primary may be higher. In other words, one cannot treat the energy loss spectrum as well as the muons' behaviors themselves in consistent manner in \textit{the differential method} with the constant values of $v_{cut}$.

Summarized speaking, the muon loss energy spectrum obtained by \textit{ the differential method}, probably, one cannot construct the reliable energy loss spectrum by which Cherenkov lights production spectrum are obtained.

In Figures 54 to 57,  we give the average transition curves for Cherenkov lights as the functions of the observation points over 1 kilometers. Each sampling number is 500 except for $10^{15}$ eV (100 sampling). The uncertainty bars denote the range within which 68\% of the total events are included, which correspond to the normal distribution with the average values and standard deviations obtained in these distributions. It is easily understood from these figures that the uncertainties of Cherenkov lights at each depth distribute over one order of magnitude except the case for $10^{12}$ eV and it is almost pretty difficult for us to decide the muon energies from the corresponding Cherenkov lights. Namely, this means that it is almost difficult for us to construct the corresponding neutrino energies directly from the muon energies.

Instead, an alternative way is suggested to decide the neutrino energy spectrum and muon energy spectrum simultaneously, via Chrenkov lights Production spectrum.

Finally, we comment on the influence of LPM effect on Cherenkov light production spectrum. There are two different the LPM effect. The Influence of the LPM effect on the muon itself can be neglected up to $10^{21}$ eV [14],[15]. The LPM effect on electron and photons [16],[17],[18],[19],[20],[21] is surely important factor in high energy neutrino astroparticle physics, particularly via electromagnetic cascade shower. The LPM effect becomes effective in the level of the cross sections at $\sim 10^{15}$ eV in water. However, it become effective at $\sim 10^{17}$ eV or more in the level of the electromagnetic cascade shower in water.

In the subsequent paper, we would present the alternative way for the construction of the neutrino spectrum  via Cherenkov lights production spectrum.

\newpage
\appendix
\setcounter{figure}{0}
\section{}
\label{App}
 If the differential cross section for bremsstrahlung\cite{Kelner}, direct 
pair production\cite{Kokoulin} and photo nuclear interaction\cite{Borg} are denoted 
by $\sigma_{b}\left(E_{0},E_{\gamma}\right)dE_{\gamma}$, $\sigma_{d}\left(E_{0},E_{e}\right)dE_{e}$ and $\sigma_{n}\left(E_{0},E_{h}\right)dE_{h}$, 
respectively. \\
 Then, the mean free paths of the muons due to bremsstrahlung, direct electron pair production and photo nuclear interaction are given, respectively, as
\begin{equation}
\lambda_{b}\left(E_{0}\right)= \frac{1}{\frac{N}{A}\int_{E_{\gamma_{min}}}^{E_{\gamma_{max}}} \sigma_{b}\left(E_{0},E_{\gamma}\right)dE_{\gamma}}\\
\end{equation}
\begin{equation}
\lambda_{dp}\left(E_{0}\right)= \frac{1}{\frac{N}{A}\int_{E_{ep_{min}}}^{E_{ep_{max}}} \sigma_{dp}\left(E_{0},E_{ep}\right)dE_{ep}}
\end{equation}
\begin{equation}
\lambda_{n}\left(E_{0}\right)= \frac{1}{\frac{N}{A}\int_{E_{h_{min}}}^{E_{h_{max}}} \sigma_{n}\left(E_{0},E_{h}\right)dE_{h}}
\end{equation}
 Also, the resultant mean free path for there radiative processes are given as
\begin{equation}
\frac{1}{\lambda_{total}\left(E_{0}\right)}= \frac{1}{\lambda_{b}\left(E_{0}\right)}+\frac{1}{\lambda_{dp}\left(E_{0}\right)}+\frac{1}{\lambda_{n}\left(E_{0}\right)}
\end{equation}
The integrations for (A.1) to (A.3) are performed over kinematically allowable ranges.
In (A.1), $E_{\gamma, min}$ is taken to be satisfied with such a condition that $E_{\gamma, min}/E_{0}$ sufficiently smaller than $E_{min}/E_{0}$, where $E_{0}$ and $E_{min}$ denote the primary energy of the muon and the minimum energy of the muon for observation.  $E_{min}$ is taken as 1 Gev throughout present paper. 
In Figure A.1, we give the mean free paths for bremsstrahlung, direct electron pair production and photonuclear interaction are given as the function of the primary energy. 
\begin{figure}[h]
\begin{center}
\resizebox{0.5\textwidth}{!}{%
  \includegraphics{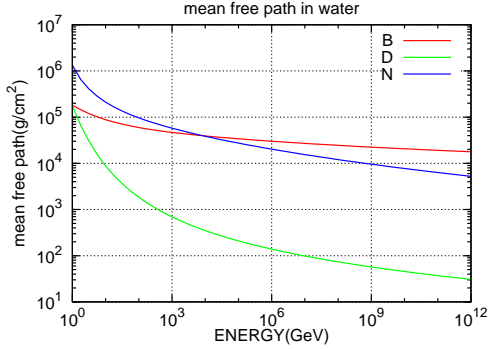}}
\caption{Mean free paths for bremsstrahlung, direct electron pair production, photo nuclear interaction and their total processes}
\label{fig:MFP}
\end{center}
\end{figure}\\
 The most important procedures in our Monte Carlo method are only two procedures.\\
The first one is where interaction occurs.\\
For the bremsstrahlung,\\
 The traversed distance for the interaction is determined with the use of $\xi $, uniform random number between (0,1), as follows.
\begin{equation}
\Delta t = -\lambda_{b}\left(E_{0}\right)log\xi
\end{equation}
Similarly, for direct electron pair production,
\begin{equation}
\Delta t = -\lambda_{dp}\left(E_{0}\right)log\xi
\end{equation}
Similarly, for the photo nuclear interaction,
\begin{equation}
\Delta t = -\lambda_{h}\left(E_{0}\right)log\xi
\end{equation}
 In our Monte Carlo simulation, the energy loss due to each stochastic  processes are sampled by the following equations with the use of $\xi$, the uniform random number, between $\left(0,1\right).$\\
 For Bremsstrahlung
\begin{equation}
\label{xi_b}
\xi = \frac{\int_{E_{\gamma_{min}}}^{E_{\gamma}} \sigma_{b}\left(E_{0},E_{\gamma}\right)dE_{\gamma}}{\int_{E_{\gamma_{min}}}^{E_{\gamma_{max}}} \sigma_{b}\left(E_{0},E_{\gamma}\right)dE_{\gamma}}
\end{equation}
 From (\ref{xi_b}),
\begin{equation}
E_{\gamma} = F_{b}\left(E_{0},\xi\right)
\end{equation}
 For direct electron pair production
\begin{equation}
\label{xi_d}
\xi = \frac{\int_{E_{ep_{min}}}^{E_{ep}} \sigma_{dp}\left(E_{0},E_{ep}\right)dE_{ep}}{\int_{E_{ep_{min}}}^{E_{ep_{max}}} \sigma_{dp}\left(E_{0},E_{ep}\right)dE_{ep}}
\end{equation}
 From (\ref{xi_d}),
\begin{equation}
E_{e} = F_{d}\left(E_{0},\xi\right)
\end{equation}
 For photo nuclear interaction
\begin{equation}
\label{xi_n}
\xi = \frac{\int_{E_{h_{min}}}^{E_{h}} \sigma_{n}\left(E_{0},E_{h}\right)dE_{h}}{\int_{E_{h_{min}}}^{E_{h_{max}}} \sigma_{n}\left(E_{0},E_{h}\right)dE_{h}}
\end{equation}
 From (\ref{xi_n}),
\begin{equation}
E_{h} = F_{h}\left(E_{0},\xi\right)
\end{equation}
 For the discrimination among stochastic processes in our Monte Carlo simulation let us introduce the following equations.
\begin{equation}
\xi_{a}\left(E_{0}\right) = \frac{1/\lambda_{b}\left(E_{0}\right)}{1/\lambda_{b}\left(E_{0}\right)+1/\lambda_{dp}\left(E_{0}\right)+1/\lambda_{n}\left(E_{0}\right)}
\end{equation}
\begin{equation}
\xi_{b}\left(E_{0}\right) = \frac{1/\lambda_{b}\left(E_{0}\right)+1/\lambda_{dp}\left(E_{0}\right)}{1/\lambda_{b}\left(E_{0}\right)+1/\lambda_{dp}\left(E_{0}\right)+1/\lambda_{n}\left(E_{0}\right)}
\end{equation}
 In Figure \ref{Flow}, we give a flow chart for our Monte Carlo simulation.
\begin{figure}[h]
\begin{center}
\resizebox{1.5\textwidth}{!}{%
\includegraphics{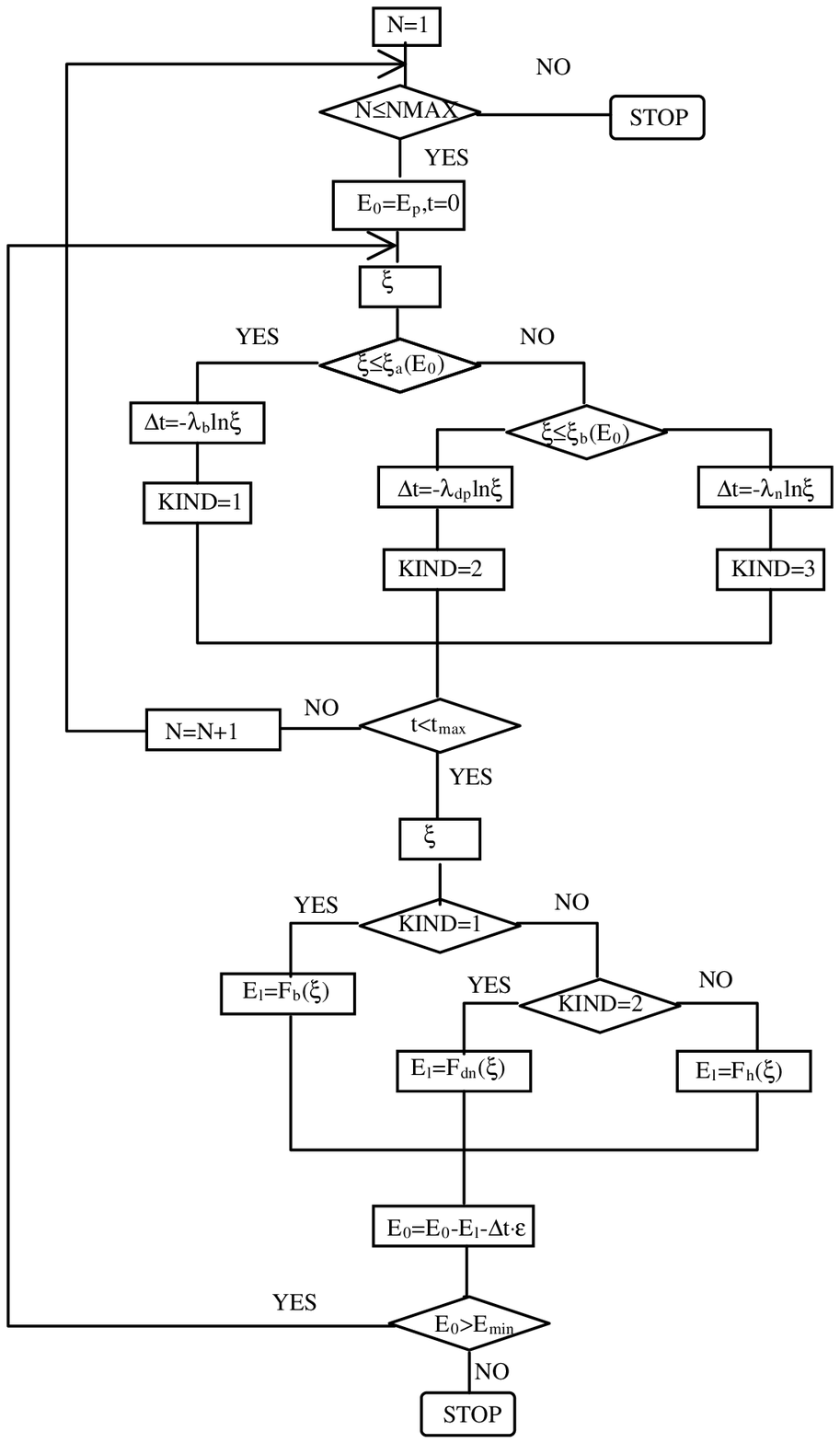}}
\caption{}
\label{Flow} 
\end{center}
\end{figure}
The validity of our Monte Carlo procedure by \textit{the integral method} should be carefully examined in the two methods which are independent of each other. Namely, The first one is the comparison of our procedure with the analytical theory which is explained in the present Appendix and the other is the comparison of our procedure (\textit{the integral method}) with the different kind of Monte Carlo procedure (\textit{the differential method}) which is mentioned in the text.

Particularly, it is the best that the results obtained by a Monte Carlo procedure are checked by the procedures which is methodologically independent of the Monte Carlo procedure, reaching the same results.

In Figure A.3, the average energies of the muons are given as the function of the depths under the preposition of muon energy spectrum at sea level with indices 2, 3 and 4, obtained by \textit{the integral method} [12, 22] and they are compared with results obtained by the analytical theory based on the Nishimura-Kamata formalism in the cascade shower theory and the agreement between them [23] are very well, which surely guarantee the validity of our \textit{integral method}.


\begin{figure}[h]
\begin{center}
\resizebox{0.5\textwidth}{!}{%
\includegraphics{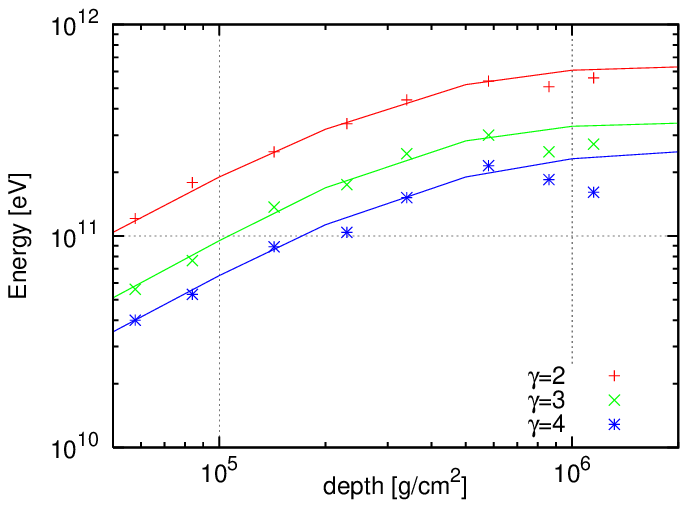}}
\caption{}
\label{fig}
\end{center}
\end{figure}

\newpage
\bibliography{<your-bib-database>}

\end{document}